\documentclass[usenatbib]{mn2e}
\usepackage{graphicx}

\title[Results from WASP0 III: Planet Hunting in the Draco Field]
{Results from the Wide Angle Search for Planets Prototype (WASP0) III:
Planet Hunting in the Draco Field}
\author[S. R. Kane et al.]{Stephen R. Kane$^1$, Andrew Collier Cameron$^1$,
Keith Horne$^1$, David James$^{2,3}$,
\newauthor T. A. Lister$^1$, Don L. Pollacco$^4$, Rachel A. Street$^4$,
Yiannis Tsapras$^5$\\
$^1$School of Physics \& Astronomy, University of St Andrews, North Haugh,
St Andrews, Fife KY16 9SS, Scotland\\
$^2$Department of Physics \& Astronomy, Vanderbilt University, Nashville,
TN 37235, USA\\
$^3$Laboratoire d'Astrophysique, Observatoire de Grenoble, BP 53, F-38041,
Grenoble, Cedex 9, France\\
$^4$School of Mathematics and Physics, Queen's University, Belfast,
University Road, Belfast, BT7 1NN, Northern Ireland\\
$^5$School of Mathematical Sciences, Queen Mary University of London,
Mile End Road, London, E1 4NS, UK}

\begin{document}

\maketitle

\begin{abstract}

The Wide Angle Search for Planets prototype (WASP0) is a wide-field
instrument used to search for extra-solar planets via the transit method.
Here we present the results of a monitoring program which targeted a
9-degree field in Draco. WASP0 monitored 35000 field stars for two
consecutive months. Analysis of the lightcurves resulted in the detection
of 11 multi-transit candidates and 3 single-transit candidates, two of
which we recommend for further follow-up. Monte-Carlo simulations matching
the observing parameters estimate the expected number of transit
candidates from this survey. A comparison of the expected number with the
number of candidates detected is used to discuss limits on planetary
companions to field stars.

\end{abstract}


\section{Introduction}

The field of extra-solar planet detection has seen rapid expansion over
recent years, both in the number of teams working in the field and in the
number of planets detected. This recent expansion has been partly due
to the realized potential of the transit method and the ability of
relatively cheap instruments as effective tools in searching for
transiting extra-solar planets. This was made apparent by the first
observation of the transiting planet HD 209458b \citep{cha00,hen00}. Since
then instruments such as STARE \citep{bro99}, Vulcan \citep{bor01}, and
HAT \citep{bak02} have greatly contributed to the sky coverage in the
search for planetary transits.
This has led to other successful detections of transiting planets, such as
OGLE-TR-56b \citep{kon03} and TrES-1 \citep{alo04}.
In addition to wide-field searches for transiting planets, a number of
narrow-field transit surveys of stellar clusters have been undertaken by
several groups (eg., \citet{moc02,str02}). These narrow-field surveys
tend to concentrate on open clusters which are a rich source of young,
metal-rich stars.

The reason that the transit method of extra-solar planet detection has
become so popular is due to the radial velocity surveys discovering a
relatively high number of ``hot Jupiters'' orbiting solar-type stars. In
fact, 0.5\%--1\% of Sun-like stars in the solar neighbourhood have been
found to harbour a Jupiter-mass companion in a 0.05 AU (3--5 day) orbit
\citep{lin03}. It is reasonable to assume that the orbital plane of these
short-period planets are randomly oriented, which means that approximately
10\% of these planets will transit the face of their parent star as seen by
an observer. Thus, the transit method is favoured considering the
conclusion that close to 1 in 1000 solar-type stars will produce detectable
transits due to an extra-solar planet. Since this transit method clearly
favours large planets orbiting their parent stars at small orbital radii,
a large sample of stars must be monitored in order to detect statistically
meaningful numbers of transiting planets.

The Wide Angle Search for Planets prototype (hereafter WASP0) is an
inexpensive wide-field instrument \citep{kan04} developed as a precursor
to SuperWASP \citep{str03}, a more advanced instrument that has recently
been constructed on La Palma, Canary Islands. The primary science goal of
both instruments is the detection of transiting extra-solar planets. The
first WASP0 observing run was undertaken on La Palma, where observations
concentrated on a field in Draco which was regularly monitored for two
months. In order to monitor sufficient numbers of stars for successful
planetary transit detection, a wide field needs to be combined with
reasonably crowded star fields. The Draco field was chosen due to its
circum-polar location combined with a relatively high density of field
stars. The high sampling rate and duration of monitoring of this field
results in a high sensitivity to transiting short-period planets.

We present the results from WASP0 monitoring of the Draco field for
transiting extra-solar planets. In sections 2 and 3 we describe the
observations and the data reduction methods used to achieve the millimag
accuracy required. Section 4 presents results from Monte-Carlo simulations
performed which closely match the data from observations of the Draco
field. These simulations are used to test the transit detection algorithm
described and expected numbers of detectable transiting extra-solar
planets are predicted for the data. Section 5 then presents the results of
our transit search, including single and multiple-transit candidates.
Finally, in section 6 we calculate the resulting limits on planetary
companions around field stars and discuss various methods of optimising
future transit surveys.

\section{Observations}

The WASP0 instrument is a wide-field (9-degree) 6.3cm aperture F/2.8 Nikon
camera lens, Apogee 10 CCD detector (2K $\times$ 2K chip, 16-arcsec
pixels) which was built by Don Pollacco at Queen's University, Belfast.
Calibration frames were used to measure the gain and readout noise of the
chip and were found to be 15.44 e$^-$/ADU and 1.38 ADU respectively.
Images from the camera are digitised with 14-bit precision giving a data
range of 0--16383 ADUs. The instrument uses a clear filter which has a
slightly higher red transmission than blue.

\begin{figure}
  \includegraphics[width=8.2cm]{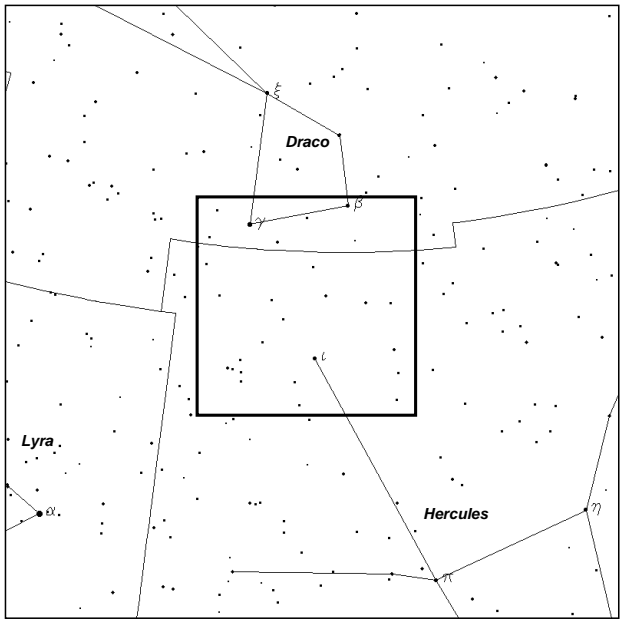}
  \caption{The observed $9^\circ \times 9^\circ$ field (shown as a square
in the center of the figure) which lies in both Draco and Hercules.}
\end{figure}

WASP0 has had two successful observing runs at two separate sites. The
first observing run was undertaken on La Palma, Canary Islands during
2000 June 20 -- 2000 August 20. The second observing run took place at
Kryoneri, Greece between 2001 October -- 2002 May. During the La Palma
run, WASP0 was mounted piggy-back on a commercial 8-inch Celestron
telescope with a German equatorial mount. These observations concentrated
on a field in Draco which was regularly monitored for two months. The
location and size of the field is shown in Figure 1. The field center was
located at RA. $17^{\mathrm{h}} 40^{\mathrm{m}} 00^{\mathrm{s}}$ and Dec.
47\degr 55\arcmin 00\arcsec. The observations were interrupted on four
occasions when a planetary transit of HD 209458 was predicted. On those
nights, a large percentage of time was devoted to observing the HD 209458
field in Pegasus.

Exposure times generally alternated between 20s and 120s to extend the
dynamic range so that brighter stars saturated in the longer exposure
would be unsaturated in the shorter exposure. However, it was later found
that only about 40 additional bright stars were gained by including the
20s frames which at the same time added significant noise to the overall
rms of the data. Hence it was decided to only include the 120s frames in
the analysis. Further details regarding the observations are described in
\citet{kan04}.

\section{Data Reduction}

To reduce the large WASP0 dataset, a data reduction reduction pipeline was
developed with a high degree of automation. A major challenge for
wide-field transit surveys is to produce accurate photometry from the
images. This is particularly difficult for wide-fields since there are
many spatially-dependent aspects which are normally assumed to be constant
across the frame, for example the airmass and heliocentric time correction.
These problems have been largely solved by implementing a flux-weighted
astrometric fit which uses both the Tycho-2 \citep{hog00} and USNO-B
\citep{mon03} catalogues. 

Rather than fit the spatially-variable point-spread function (PSF) shape of
the stellar images, we used weighted aperture photometry to compute the
flux in a circular aperture of tunable radius centred on the predicted
positions of all objects in the catalogue. The weights assigned to pixels
lying partially outside the aperture are computed using a Fermi-Dirac-like
function which are then renormalised to ensure that the effective area of
the aperture is $\pi r^2$ where $r$ is the aperture radius in pixels. For
the WASP0 data, fluxes were computed using apertures of radii 1.5, 2.5,
and 3.5 pixels. These fluxes shall be referred to as $F_1$, $F_2$, and
$F_3$ respectively.

The most serious issues arise from vignetting and barrel distortion,
produced by the camera optics, which alter the position and shape of
stellar profiles. In particular, this can result in serious blending
effects for stars which neighbour significantly distorted stellar profiles.
It has been shown by \citet{bro03} and \citet{tor04} that blending can have
a significant effect on transit searches.

\begin{figure}
  \includegraphics[width=8.2cm]{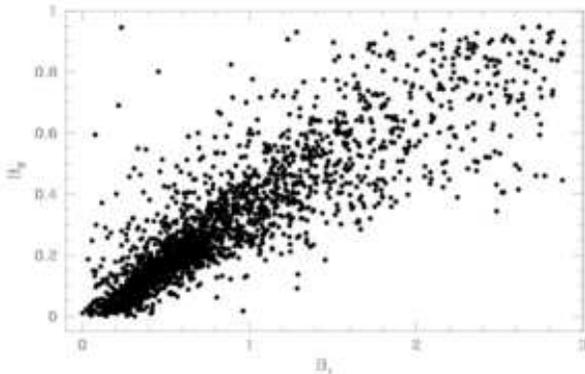}
  \caption{Plot of the two ratios used to distinguish blended from
unblended stars. The unblended stars tend to form a tight loucus of stars
close to the origin.}
\end{figure}

In order to identify stars significantly effected by blending, we compute
a blending index using the flux measurements from the three different
aperture radii. First, we consider the two ratios:
\begin{equation}
  B_1 = \frac{F_3 - F_1}{F_1}
\end{equation}
and
\begin{equation}
  B_2 = \frac{F_3 - F_2}{F_2}
\end{equation}
These ratios are plotted in Figure 2 for a typical frame. The purpose of
this blend ratio is to provide information about the flux in the core of
the stellar profile compared with the flux in the wings.
The blend ratio for blended sources tends to increase with a larger
aperture size as more flux from neigbouring sources contributes to the
total measured flux. The main body of stars in the main locus close to the
origin are generally unblended objects whose core-wing ratios are
determined by the colour of the star and the chromatic aberration in the
lens. The blended stars tend to fall above or below the main locus or fall
into a fan which extends beyond the main locus. Direct inspection of the
images showed that objects with $(F_2-F_1)/F_1$ greater than about the
80th percentile value were invariably found to be either M supergiants or
blends, if they lay close to a linear fit to the main locus. Hence, we
chose to flag all objects close to the line defined by the main locus with
$(F_2-F_1)/F_1$ greater than the 80th percentile as likely blends.
This method can therefore be used to approximately classify stars as
blended or unblended, allowing the blended stars to be excluded from the
analysis.

Post-photometry calibrations are applied to remove time-dependent and
position-dependent trends from the data. The post-photometry calibration
code constructs a theoretical model which is then subtracted from the
data leaving residual lightcurves. The residuals are then iteratively
fitted to calibrate and remove systematic correlations in the data. Rms
accuracy versus magnitude plots are available to evaluate the improvement
by applying the model. The de-trended lightcurves are then further
analysed for periodic variability including transit signatures. The
reduction of the WASP0 data is described in more detail in \citet{kan04}.

\section{Monte-Carlo Simulations}

To estimate how many transit events we expect to see in our data, we
performed Monte-Carlo simulations which inject transits into fake data
and test the capabilities of the transit detection algorithm.

\subsection{Simulated Transit Lightcurves}

The probability of an observable planetary transit occurring depends upon
the inclination of the planet's orbital plane $i$ satisfying $a \cos i
\leq R_\star + R_p$ where $R_\star$ and $R_p$ are the radii of the star
and planet respectively and $a$ is the semi-major axis of the planet's
orbital radius. The transit probability is then given by $(R_\star + R_p)
/ a$. As far as the probability is concerned, the size of the planet is of
little consequence and depends mostly upon the size of the parent star and
orbital radius.

\begin{figure*}
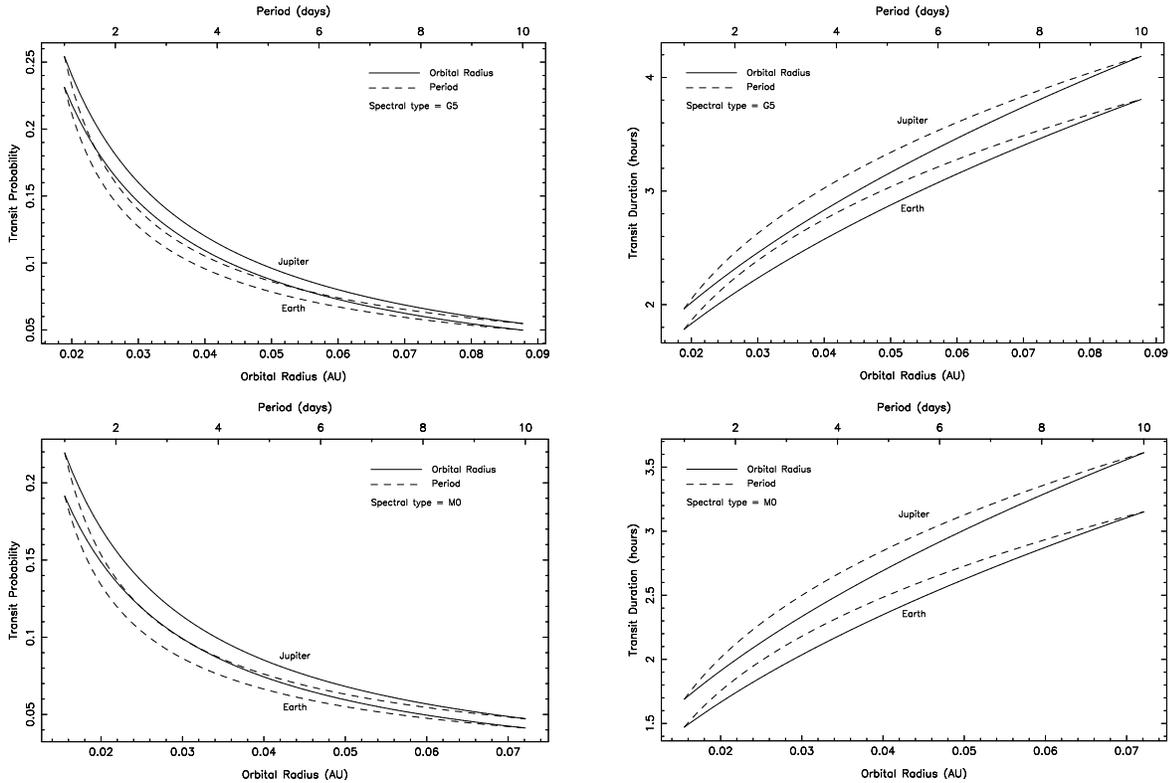

  \begin{center}
    \begin{tabular}{cc}
      \includegraphics[angle=270,width=7.2cm]{figure03a.ps} &
      \hspace{0.5cm}
      \includegraphics[angle=270,width=7.2cm]{figure03b.ps} \\
      \includegraphics[angle=270,width=7.2cm]{figure03c.ps} &
      \hspace{0.5cm}
      \includegraphics[angle=270,width=7.2cm]{figure03d.ps} \\
    \end{tabular}
  \end{center}
  \caption{The geometric transit probability (left) and transit duration
(right) for planets of Earth and Jupiter radii. The period range has been
chosen to match the sensitivity of our survey to multiple transits.}
\end{figure*}

The Draco field was monitored by WASP0 on 29 nights over a period of 35
nights. The Monte-Carlo simulations and transit search has therefore been
optimised for an orbital period range of 1--10 days since this is the
period range that will most likely yield multiple transiting events for
our survey. Shown in Figure 3 are probability and duration plots for Earth
and Jupiter radius planets orbiting G5 and M0 main sequence stars. It can
be seen from the duration plots that the duration ranges between 2--4
hours for a solar-type star and is a small function of planet radius.

The planetary radius does however matter a great deal for the transit to be
detectable, since the fractional transit depth is
\begin{equation}
  \frac{\Delta F}{F_0} \approx \left( \frac{R_p}{R_\star} \right)^2
\end{equation}
where $F_0$ is the baseline flux of the star and $\Delta F$ maximum change
in flux due to a transit. Shown in Figure 4 are model transits and
simulated data for a range of magnitudes and spectral types assuming a
single transit by a Jupiter-radius planet and data binning of 10 minutes.
The noise model used takes into account detector characteristics as well
as photon statistics and takes the form
\begin{equation}
  \sigma^2 = \sigma_0^2 + \frac{(f_\star + f_\mathrm{sky}) \Delta t}{G}
\end{equation}
where $\sigma_0$ and $G$ are the CCD readout noise (ADU) and gain
(e$^-$/ADU) respectively, $f_\star$ and $f_\mathrm{sky}$ are the
star and sky fluxes respectively, and $\Delta t$ is the exposure time.
The plot windows have been normalised to a width of $3 R_\odot / R_\star$,
equivalent to the projected path of the planet as it crosses the stellar
disk. The depth of the lightcurves is shown to be the same in each case
because, although the depth actually varies a great deal, the plots aim
to compare the width and shape of the lightcurves and the accuracy of the
photometric measurements. Transits of late-type stars are of shorter
duration but are considerably deeper and so produce a much higher
signal-to-noise (S/N) during the transit. For the WASP0 detector, transits
around solar-type stars fainter than 12th magnitude become undetectable
although folding data on the orbital period will improve the S/N.

\begin{figure}
  \includegraphics[angle=270,width=8.2cm]{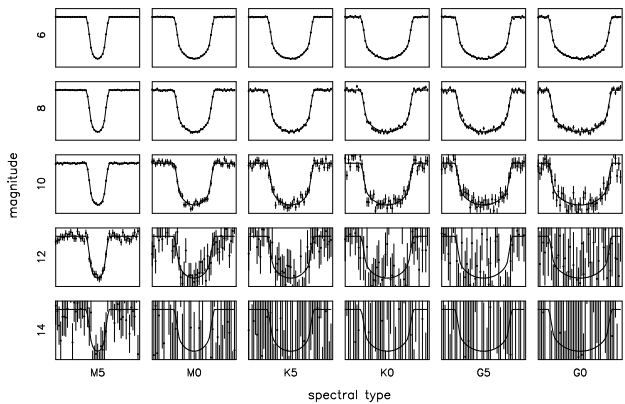}
  \caption{Model transits lightcurves overlaid with simulated data from
the WASP0 instrument for a range of spectral types and magnitudes. This
assumes the transiting planet is of approximately Jupiter radius.}
\end{figure}

\begin{figure}
  \includegraphics[angle=270,width=8.2cm]{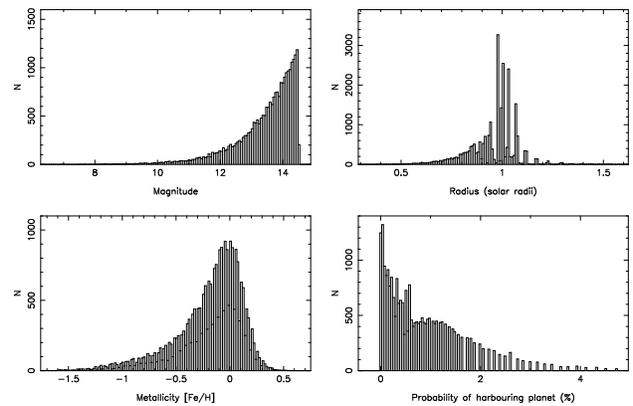}
  \caption{Properties of stars included in the simulation sample, as
derived from the Besan\c{c}on model. Distributions include the magnitude
(top-left), the stellar radii (top-right), the metallicities (bottem-left),
and the probability of harbouring a planet (bottom-right).}
\end{figure}

By applying the noise model shown in equation 4 to the WASP0 detector,
simulated lightcurves were generated using the sky values and epochs from
the observations. A Besan\c{c}on model \citep{rob03} tailored to the Draco
field observations created a distribution of magnitudes, colours, and
metallicities from which stellar parameters were derived. Stellar radii
for main sequence stars were calculated directly from the colours provided
and giant stars were excluded from the analysis. By utilising the
planet-metallicity correlation presented in \citet{fis05} which relates
stellar metallicity to planetary abundance, the probability of each star
harbouring a planet was also computed. Since we are only considering hot
Jupiters with a period range of 1--10 days, the resulting probabilities
were appropriately weighted by assuming that the periods are approximately
uniform in log space \citep{tab02}. Figure 5 shows the histograms of the
resulting sample of stars used for the simulations. The stellar population
in the Draco field is largely comprised of G dwarf stars with approximately
solar metallicity. The power law nature of the \citet{fis05} correlation
tends to dramatically increase the number of stars with low
planet-harbouring probability for a typical metallicity distribution. In
particular, stars with metallicities less than $\sim -0.5$ were have
essentially zero probability of harbouring a planet, thus resulting in the
sharp rise shown in Figure 5.

The existence of a hot Jupiter companion for each star in the sample was
randomly determined based on the probability for that star hosting a
planet. In cases where a planet was deduced to exist, the planetary
radius and the period and inclination of the planetary orbit were randomly
generated. Planetary radii were allowed to vary between $0.5 R_J < R_p <
1.5 R_J$ where $R_J$ is equivalent to one Jupiter radius. These planetary
and stellar radii result in a range of transit depths from 0.1\% to 25\%.
The period was allowed to vary uniformly in log space between 1 and 10 days.
In the case of stars which contain a planet with a favourable orbital
orientation, the reduction in flux due to a planetary transit was inserted
into the data if the star was observed at those epochs. This simulation
was run multiple times and a statistical analysis of the expected transit
parameters was performed.

Running this simulation many times yielded an average of $\sim 17$ stars
with injected transits up to the magnitude limit of 14.5. The distribution
of the stars with transits naturally scales with the magnitude
distribution shown in Figure 5.
The number of cases with transits occurring during observations of
the star was found to be $\sim$ 94\% including single transits. The number
of multiple transits observed was found to be $\sim$ 86\%.
The probability of observing multiple transits during the
course of our observations was calculated for individual periods between 1
and 10 days. This period scan was used to produce the plot shown in Figure
6 which clearly shows the major reduction in probability for periods close
to an integer number of days. The figure also shows that almost all
transiting planets with periods between 1 and 4 days will produce multiple
transits in the data.

\begin{figure}
  \includegraphics[angle=270,width=8.2cm]{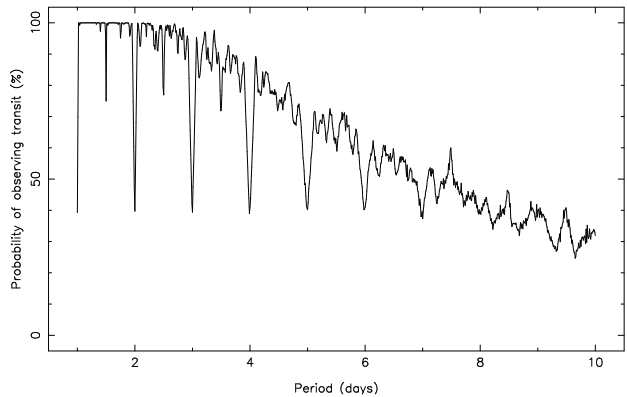}
  \caption{Probability of observing at least two transits of a star with
injected transits during Draco observations as a function of period.}
\end{figure}

Although Figure 6 demonstrates the high probability for our data of
observing stars with at least two transits, it does not take into
account the detectablility of the transit. This restriction is limited by
the S/N of the transit and the rms of the associated lightcurve. By
running the simulation just once, an entire dataset of lightcurves was
generated including injected transits. In order to test the detectability
aspect, the lightcurves generated by the simulation were used as input for
the transit detection algorithm.

\subsection{Transit Detection Algorithm}

The ability to efficiently detect the signature of a planetary transit in
thousands of lightcurves has been a major challange for the transit
survey teams. Automation of transit detection has hence become vigorously
studied and several methods have been suggested (eg.,
\citet*{def01,doy00,kov02}). Two of the important issues for such methods
are the reduction of computational time to a reasonable value and the
optimisation of the model to avoid false positive detections. The method
used here is a matched-filter algorithm which generates model transit
lightcurves for a selected range of transit parameters and then fits them
to the stellar lightcurves.

The transit model used for fitting the lightcurves is a truncated cosine
approximation with four parameters: period, duration, depth, and the time
of transit midpoint. The search first performs a period sweep, optimising
the depth and centroid, but holding fixed the duration. The advantage of
fixing the duration to a reasonable value and then scanning for multiple
transits is that it dramatically reduces the number of false positive
detections by avoiding single-dip events. The search is refined by
optimising the duration for those stars which are fitted significantly
better by the transit model compared with a constant lightcurve model. A
transit S/N statistic is calculated for each lightcurve based on the
resulting reduced $\chi^2$ and $\Delta \chi^2$ as follows:
\begin{equation}
  S_W^2 = \frac{\Delta \chi^2}{\chi_{\mathrm{min}}^2 / (N - f)}
\end{equation}
where $N$ is the number of data points, $f$ is the number of free
parameters, and $\Delta \chi^2$ is given by
\begin{displaymath}
  \Delta \chi^2 = \chi_{\mathrm{constant}}^2 - \chi_{\mathrm{transit}}^2.
\end{displaymath}
The error bars in the individual lightcurves are rescaled by a factor to
force $\chi_{\mathrm{min}}^2 / (N - f) = 1$ for each lightcurve. The $S_W$
statistic shown in equation 5 is used consistently throughout the transit
detection algorithm, including the first pass in which the transit
duration is fixed. By ranking the stars in order of decreasing transit
S/N, this becomes an effective method to sift transit candidates from the
data.

\subsection{Expected Numbers}

The simulated lightcurves produced as described in section 4.1 were then
analysed using the transit detection algorithm. This exercise is useful
for two reasons: to verify the detectability of transits with low S/N and
to test the robustness of the detection algorithm.

Around 35000 simulated lightcurves with injected transits were inserted
into the transit detection algorithm. A period sweep of 1.1 to 10 days
was performed with a fixed duration of 3 hours. This is a fairly
computationally expensive exercise for such a large number of stars over
such a relatively broad period range. A plot demonstrating the result of
this experiment is shown in Figure 7, where the stars with injected
transits that transit during observations are shown as 5-pointed stars.

\begin{figure}
  \includegraphics[angle=270,width=8.2cm]{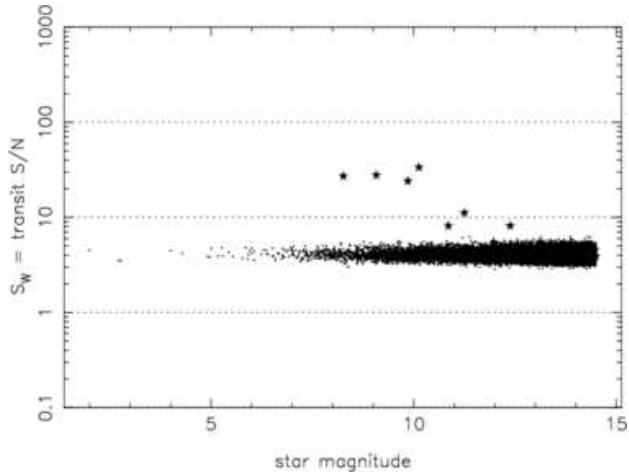}
  \caption{Results of passing the simulated data through the transit
detection algorithm. The transit S/N, which measures the
``goodness-of-fit'' of each lightcurve to a transit model, plotted against
magnitude. The stars with injected transits are shown on the diagram as
5-pointed stars.}
\end{figure}

Figure 7 shows that some of the stars with injected transits have been
successfully separated from the bulk of the lightcurves with the majority
having a transit S/N $> 10$. This also demonstrates the S/N limitations
of the data as the number of successful transit extractions decreases
significantly as the magnitude limit is approached. In total, $\sim$ 20\%
of the injected transits were recovered by the detection algorithm. Thus,
the expected number of detectable multiple transits in a dataset of 35000
stars is $\sim 47 \times 86\% \times 20\% \approx 3$. Additional factors
concerning the completeness of the lightcurves and the fraction of
blended stars will later be also taken into account.

\subsection{Establishing a Candidate Criteria}

There are several parameters that can be used to establish a selection
criteria for eliminating transit mimics from the list of candidates.
These are used consistently as a robust means for producing the final
transit candidate list presented in this paper.

{\bf Depth:} One of the first transit parameters to be yielded from a
transit fit is the depth of the transit. As discussed in section 4.1,
the expected range of transit depths using typical stellar/planetary
radii is 0.2\% to 25\%. Hence, candidates with depths significantly
outside this range can be immediately rejected.

{\bf Duration--Period:} As shown in Figure 3, there is a relation
between transit period and duration that is relatively insensitive to
stellar and planetary radii. For example, a transit duration of $\sim
7$ hours would require a period much larger than 10 days, and also a
certain amount of luck since the transit probability becomes extremely
low outside of 10 days. If the period for a candidate is calculated to
be greater than 10 days then the candidate can be rejected.

{\bf Transit shape:} The shape of the transit lightcurve can be
ambiguous in some cases, but a ``V-shaped'' signature can indicate that
the candidate is in fact a grazing eclipsing binary rather than being
due to an eclipsing planet.

{\bf Colour:} Assuming that the star is on the main sequence, the colour
combined with the depth allows an approximate calculation of the
planetary radii. Clearly this will be more favourable for red stars
rather than blue stars.

{\bf Multiple transits:} Transits candidates for which only one transit
is observed are still considered to be candidates. However, since this
means that the period is highly uncertain then the candidate is treated
with a much higher degree of skeptisicm than candidates for which
multiple transits have been observed.

\section{Results}

This section presents results from the analysis of 29 nights of
monitoring the Draco field, including photometric accuracy achieved and
transit candidates due to extra-solar planets.

\subsection{Photometric Accuracy}

Achieving the photometric accuracy necessary to be sensitive to transiting
extra-solar planets is one of the many challenges that faces wide-field
transit-hunting projects such as WASP0. For WASP0, this has been overcome
using the previously described pipeline with very good results, as
demonstrated by the rms versus magnitude diagram shown in Figure 8. The
instrumental magnitudes are roughly calibrated to $V$ using the Tycho-2
$V$ magnitudes available for the measured stars. The data shown include
around 17600 stars at 137 epochs from a single night of WASP0 observations
and includes only those stars for which a measurement was obtained at
$> 90$\% of epochs. The upper curve in the diagram indicates the
theoretical noise limit for aperture photometry with the 1-$\sigma$ errors
being shown by the dashed lines either side. The lower curve indicates the
theoretical noise limit for optimal extraction using PSF fitting.

\begin{figure}
  \includegraphics[angle=270,width=8.2cm]{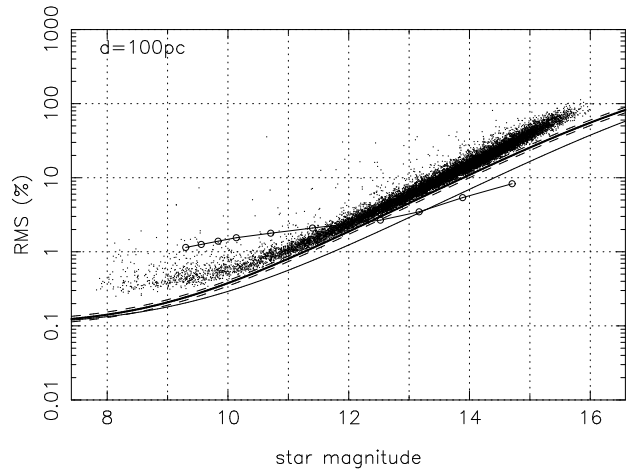}
  \caption{Photometric accuracy versus magnitude diagram from one night of
WASP0 observations, showing the rms accuracy in magnitudes in comparison
with the theoretical accuracy predicted based on the CCD noise model. The
circled line indicates the detection threshold for a Jupiter transiting
a solar-type star assuming a distance of 100 pc.}
\end{figure}

Figure 8 shows that the accuracy achieved for a single night of data is
approximately 3 mmag at the bright end. The circled line overlaid on the
diagram is for comparison with a predicted transit depth based upon a
planet of Jupiter-radius orbiting various main sequence stars. The
spectral types range from around solar at the bright end to a late-type
(M5) at the faint end. This assumes that the distance to the host star is
$d = 100 \ \mathrm{pc}$. However, since this is a comparison with the
per-data-point rms, it is lower than the sensitivity to transit detection
which is roughly given by $\mathrm{rms} / \sqrt{P_t}$ where $P_t$ is the
number of data points inside the transit.

\subsection{Planetary Transit Search}

\begin{figure*}
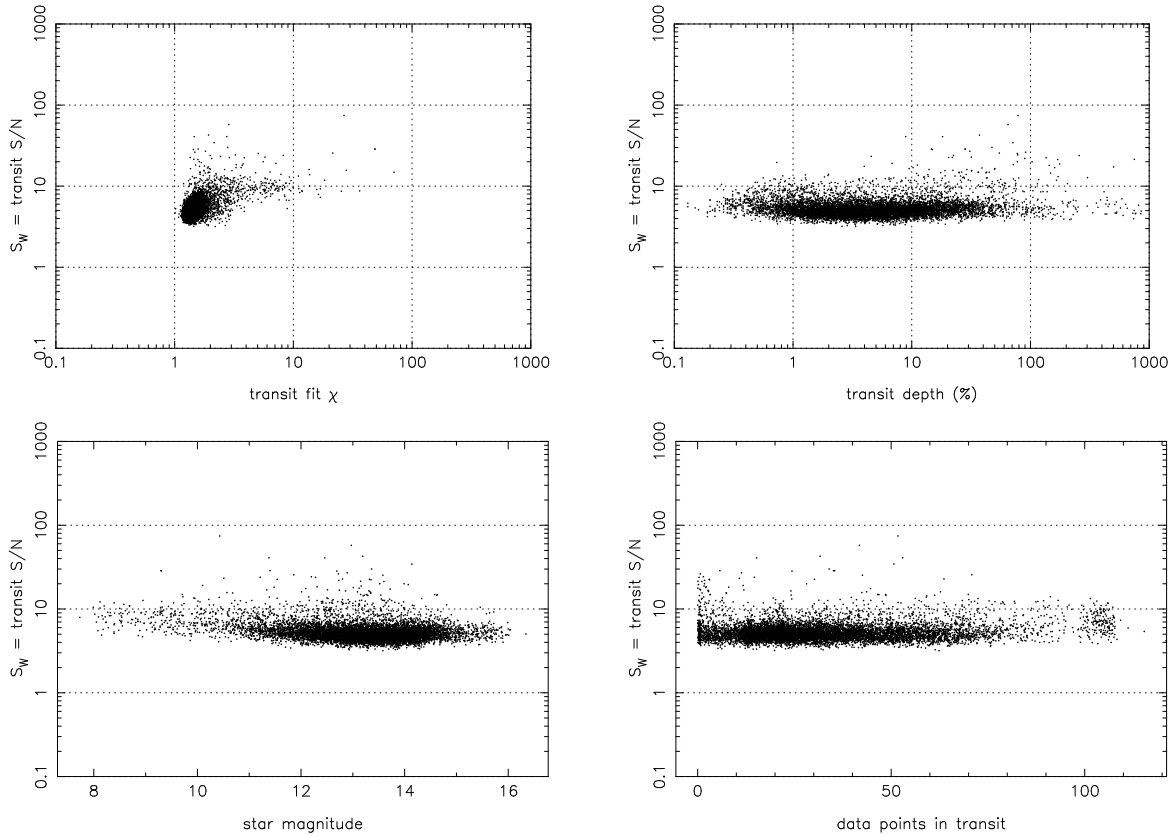

  \begin{center}
    \begin{tabular}{cc}
      \includegraphics[angle=270,width=7.2cm]{figure09a.ps} &
      \hspace{0.5cm}
      \includegraphics[angle=270,width=7.2cm]{figure09b.ps} \\
      \includegraphics[angle=270,width=7.2cm]{figure09c.ps} &
      \hspace{0.5cm}
      \includegraphics[angle=270,width=7.2cm]{figure09d.ps} \\
    \end{tabular}
  \end{center}
  \caption{Results of passing the data through the transit detection
algorithm with a period range of between 1 and 10 days with a fixed duration
of 3 hours. Each of the plots are plotted against the transit S/N which
measures the ``goodness-of-fit'' of each steller lightcurve to a transit
model.}
\end{figure*}

Before subjecting the data to the transit search algorithm, we required
that a number of conditions (cuts) be satisfied. The first cut was the
removal of blended stars, as discussed in section 3. The effect of this on
the number of stars was a reduction of $\sim 25\%$. Finally, only stars for
which data were obtained at $> 75\%$ of epochs were included, resulting in
a dataset of $\sim 14000$ stars.

The large number of stars and epochs required that the data be divided
into four magnitude bins in order for it to be managable in terms of
memory usage. Since the transit detection algorithm currently uses a grid
search rather than an amobea search of parameter space, each magnitude bin
took 24--48 hours to perform a period sweep of 1.1 to 10 days with a fixed
duration of 3 hours. The plots in Figure 9 show how the algorithm
separates the stars with a higher transit S/N from the bulk of the data.

The stars which yielded the highest transit S/N ($S_W \ga 10$) were
investigated further by performing a duration sweep of 1 to 10 hours.
Table 1 lists the best transit candidates that were extracted using this
transit detection technique. The candidates have been named WASP0-TR-$N$,
where $N$ is the sequence number.
Further information on each of the stars, including coordinates, are available
via VIZIER \citep*{och00}.

\begin{table*}
\caption{List of stars exhibiting transit-like events, where $S_W$ is the
transit S/N, $P_t$ is the number of data points obtained during transit,
and $N_t$ is the number of transits observed.}
\begin{tabular}{@{}clcccccccccc}
candidate & catalogue \# & $S_W$ & $P_t$ & mag & depth & dur     & period &
$N_t$ & colour  & $R_\star$   & $R_p$      \\
          &             &       &       &     & (\%)  & (hours) & (days) &
      & $H - K$ & ($R_\odot$) & ($R_J$)\\
WASP0-TR-01 & Tycho 3513-00814-1 & 54.32 & 25.69 & 10.51 &  4.25 & 9.54 &
9.27 & 1 & 0.069 & 0.89 & 1.83\\
WASP0-TR-02 & Tycho 3521-00445-1 & 46.98 & 43.79 & 11.38 & 10.31 & 2.33 &
1.26 & 6 & 0.047 & 1.21 & 3.89\\
WASP0-TR-03 & USNO  1350-0283522 & 41.06 & 14.93 & 12.46 & 18.64 & 3.56 &
6.35 & 2 & 0.065 & 0.93 & 4.02\\
WASP0-TR-04 & Tycho 3519-00099-1 & 31.21 &  5.33 & 11.40 & 17.26 & 6.87 &
4.69 & 1 & 0.045 & 1.23 & 5.11\\
WASP0-TR-05 & Tycho 3519-01199-1 & 20.04 & 64.54 & 12.09 &  3.61 & 5.69 &
2.35 & 2 & 0.144 & 0.64 & 1.21\\
WASP0-TR-06 & Tycho 3519-01236-1 & 19.59 &  7.55 & 11.48 & 13.18 & 8.69 &
6.03 & 1 & 0.058 & 1.05 & 3.87\\
WASP0-TR-07 & Tycho 3508-01404-1 &  9.28 & 46.99 &  9.23 &  0.62 & 2.95 &
1.66 & ? & 0.023 & 1.60 & 1.26\\
WASP0-TR-08 & USNO  1357-0286817 & 30.18 & 28.58 & 12.69 & 14.42 & 4.09 &
3.82 & 3 & 0.073 & 0.87 & 3.30\\
WASP0-TR-09 & USNO  1369-0314583 & 21.17 & 33.85 & 13.09 & 22.75 & 1.60 &
1.46 & ? & 0.083 & 0.82 & 3.91\\
WASP0-TR-10 & USNO  1357-0286796 & 17.43 & 30.10 & 12.86 &  8.97 & 4.09 &
3.82 & 2 & 0.067 & 0.91 & 2.72\\
WASP0-TR-11 & USNO  1360-0271503 & 13.59 & 49.39 & 13.05 &  7.48 & 1.93 &
1.15 & ? & 0.089 & 0.80 & 2.19\\
WASP0-TR-12 & USNO  1351-0285919 & 11.42 & 23.80 & 12.90 &  6.53 & 2.22 &
2.16 & 5 & 0.213 & 0.49 & 1.25\\
WASP0-TR-13 & USNO  1333-0303691 & 32.83 & 43.67 & 13.35 & 43.26 & 4.29 &
3.59 & 4 & 0.053 & 1.13 & 7.43\\
WASP0-TR-14 & USNO  1408-0289086 & 11.00 &  6.66 & 13.54 & 15.48 & 1.68 &
2.09 & 2 & 0.079 & 0.84 & 3.30\\
\end{tabular}
\end{table*}

It may at first seem strange to assign a period for which only one transit
is observed. However, this means that many periods can be ruled out. These
are the periods for which a transit is predicted at a time where data
points were taken and show that the predicted transit did not occur. The
best-fit period is degenerate, however, because all periods for which the
predicted transits occur at times when there are no data points are equally
likely. The periodogram for each candidate makes this clear.

The period search used in the transit detection algorithm scans the
pre-calculated period grid whilst keeping track of the best period
encountered. For a single-transit event, there will be a large number of
ties for the best fit period since the $\chi^2$ periodogram will be flat at
the best-fit value over several period ranges. The first one that occurs
in the period grid is retained as the best period. If the search is always
conducted from short to long periods, this yields the shortest period that
is consistent with the data. This may be considered a reasonably
well-defined way of specifying the period for single-transit events.

\subsection{Colour-Magnitude Diagram}

\begin{figure}
  \includegraphics[width=8.2cm]{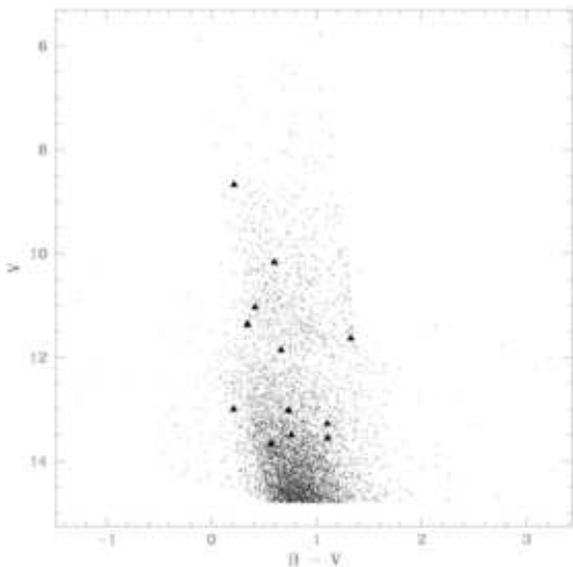}
  \caption{Colour-magnitude diagram for the Draco field with the location
of transit candidates shown as triangles.}
\end{figure}

During the data reduction process, the colours available from the
astrometric catalogues for each star are recorded in the output files
along with the measured flux. These colours are USNO-B colours or, if
available, Tycho-2 colour transformed to the USNO-B colour system. The
main difference between these two groups is the precision to which their
respective measurements have been obtained, Tycho-2 having a considerably
lower rms than the USNO-B measurements. The photometric errors present in
either catalogue are sufficiently low to allow the construction of a
rough colour-magnitude diagram. Since the Tycho-2 catalogue is 99\%
complete to $V \sim 11.0$, the source of the colour information in the
output files is a reasonable mix of the two catalogues.

Specifically, the USNO-B colours used were second epoch IIIa-J, which
approximates as $B$, and second epoch IIIa-F, which approximates as $R$.
\citet{kid04} describes a suitable linear transformation from USNO-B
filters to the more standard Landolt system. This colour transformation
is given by:
\begin{eqnarray}
  \mathrm{B: Landolt} & = & 1.097*\mathrm{USNO(B)} - 1.216 \nonumber \\
  \mathrm{R: Landolt} & = & 1.031*\mathrm{USNO(R)} - 0.417 \nonumber
\end{eqnarray}
A linear least-squares fit to the colours computed in \citet{bes90} was
used to convert from $B - R$ to $B - V$. Using this transformation, we
are able to construct an approximate colour-magnitude diagram to
investigate the relative location of the transit candidates.

The colour-magnitude diagram shown in Figure 10 appears to show no
particular colour trend for the candidates. It is expected that red
stars will have deeper transit depths than blue stars. On the other hand,
main sequence red stars will also be fainter and therefore more difficult
to perform photometry of the necessary accuracy to detect the transit.

\subsection{Stellar Radii}

\begin{figure}
  \includegraphics[angle=270,width=8.2cm]{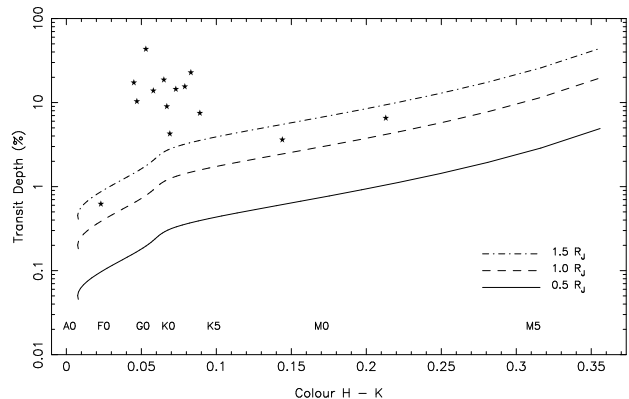}
  \caption{The transit depth for planets of radii 0.5, 1.0, and 1.5 $R_J$
as a function of colour ($H - K$). The transit candidates are shown on the
diagram as 5-pointed stars. Most of the parent stars are G--K and require
a planet radius significantly larger than Jupiter's.}
\end{figure}

For a reasonable estimate of the stellar radii of the transit candidates,
we require colours more accurate than those provided by the Tycho-2 and
USNO catalogues. Fortunately, the Two Micron All Sky Survey (2MASS)
project provides accurate colours using $J$, $H$, and $K$ filters down to
the magnitude limits of the WASP0 data. For the purposes of this study,
using $H - K$ for the colour, and hence stellar radii, determination was
the best option.

Shown in Figure 11 is a plot of the depth produced by orbiting extra-solar
planets of radii 0.5, 1.0, and 1.5 $R_J$ as a function of colour ($H - K$).
The transit candidates are shown as 5-pointed stars and are predominantly
G--K stars. Most of these stars are clustered in the top-left corner of the
diagram meaning that the transit depths exhibited by these stars require
the transiting planet to have a radius significantly larger than Jupiter's.

\subsection{Transit Candidates}

Presented here is a brief discussion for each of the candidates shown in
Table 1 and Figures 12--14. There are a number of lightcurves which
exhibit transit-like signatures but fail to satisfy the transit selection
criteria. These transit mimics are relatively common and must be
considered carefully.

\begin{figure*}
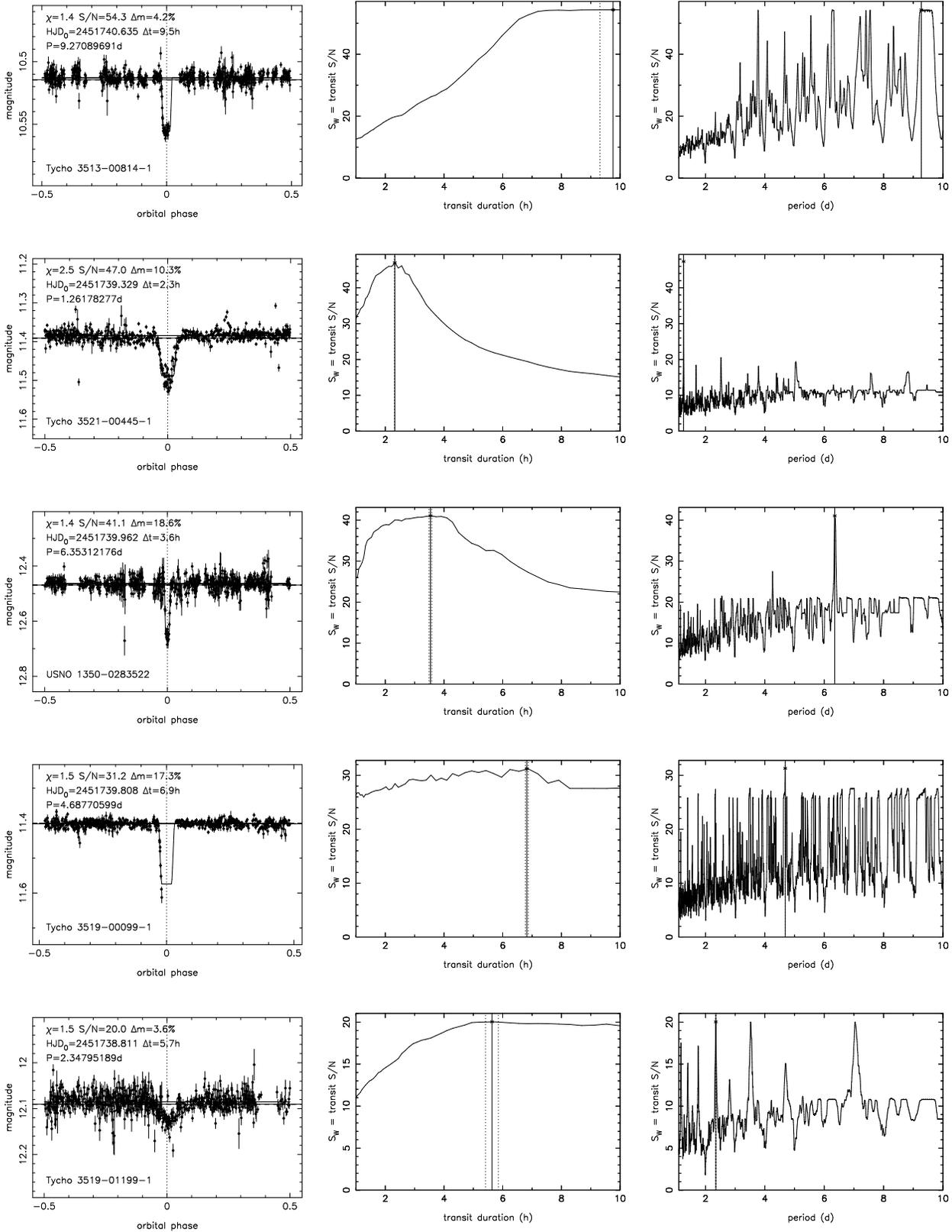

  \begin{center}
    \begin{tabular}{ccc}
      \includegraphics[angle=270,width=5.2cm]{figure12a.ps} &
      \includegraphics[angle=270,width=5.2cm]{figure12b.ps} &
      \includegraphics[angle=270,width=5.2cm]{figure12c.ps} \\
      \\
      \includegraphics[angle=270,width=5.2cm]{figure12d.ps} &
      \includegraphics[angle=270,width=5.2cm]{figure12e.ps} &
      \includegraphics[angle=270,width=5.2cm]{figure12f.ps} \\
      \\
      \includegraphics[angle=270,width=5.2cm]{figure12g.ps} &
      \includegraphics[angle=270,width=5.2cm]{figure12h.ps} &
      \includegraphics[angle=270,width=5.2cm]{figure12i.ps} \\
      \\
      \includegraphics[angle=270,width=5.2cm]{figure12j.ps} &
      \includegraphics[angle=270,width=5.2cm]{figure12k.ps} &
      \includegraphics[angle=270,width=5.2cm]{figure12l.ps} \\
      \\
      \includegraphics[angle=270,width=5.2cm]{figure12m.ps} &
      \includegraphics[angle=270,width=5.2cm]{figure12n.ps} &
      \includegraphics[angle=270,width=5.2cm]{figure12o.ps} \\
    \end{tabular}
  \end{center}
  \caption{Transit candidates 1--5 (top to bottom) from the Draco
field.}
\end{figure*}

\begin{figure*}
  \begin{center}
    \begin{tabular}{ccc}
      \includegraphics[angle=270,width=5.2cm]{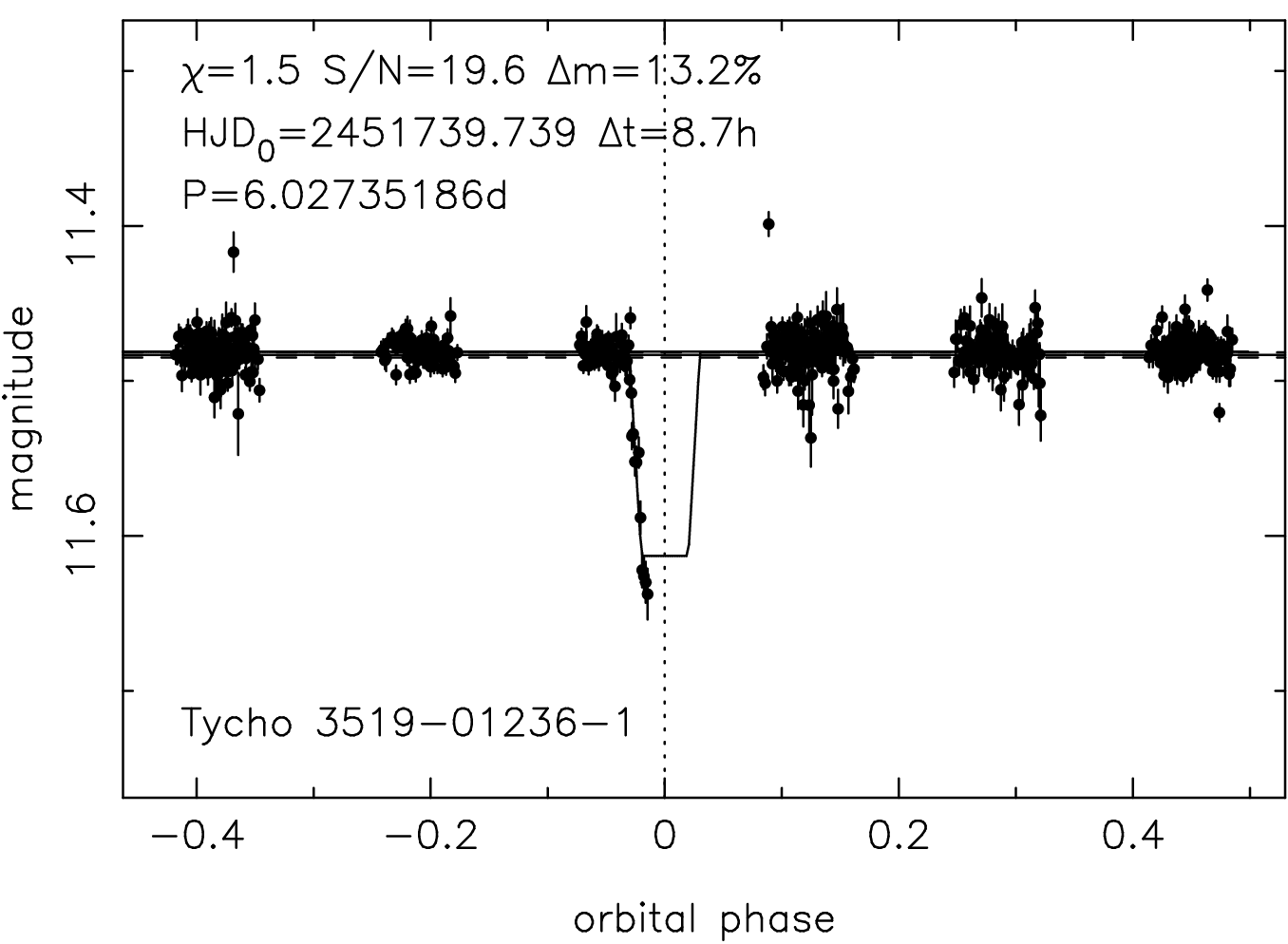} &
      \includegraphics[angle=270,width=5.2cm]{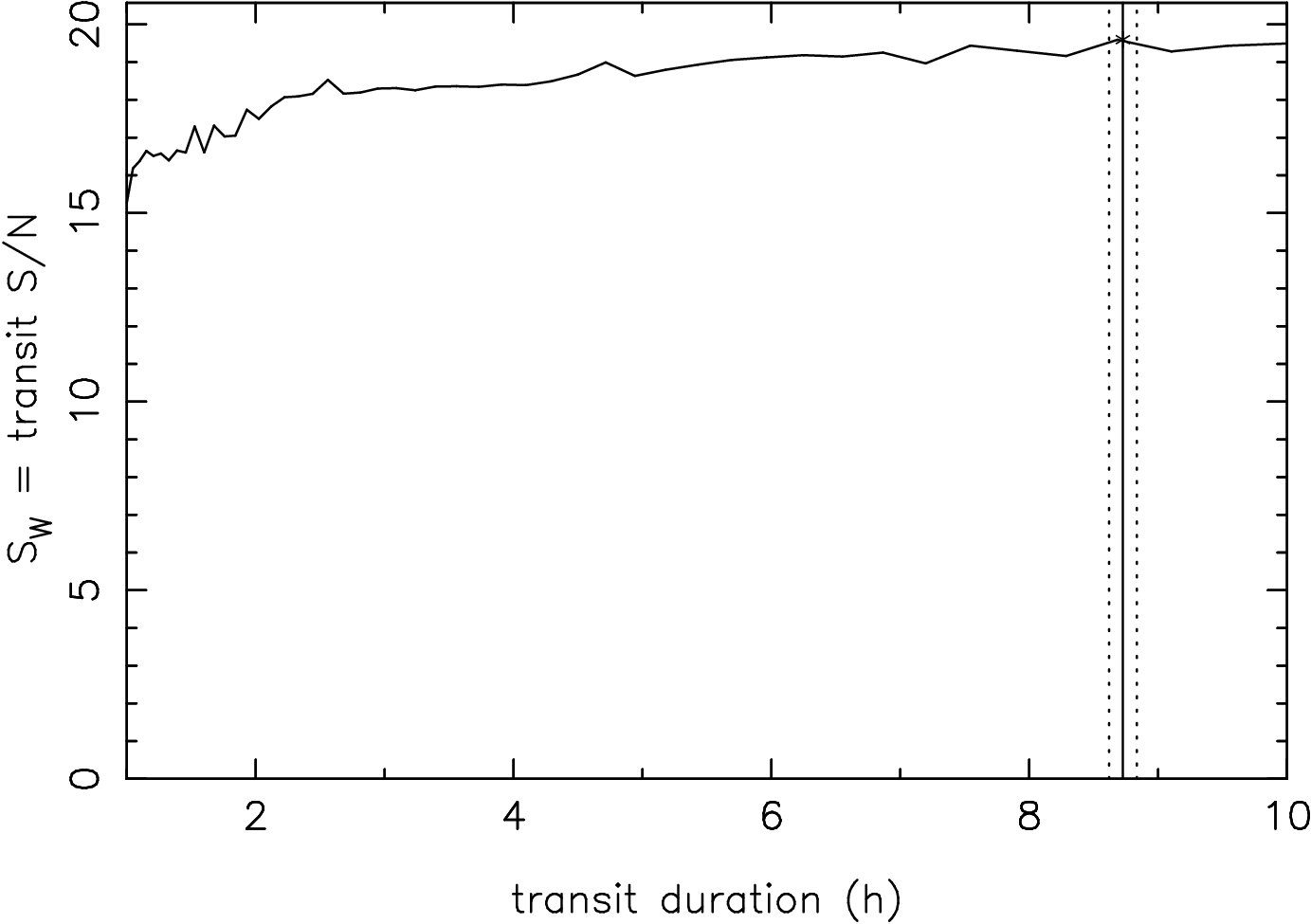} &
      \includegraphics[angle=270,width=5.2cm]{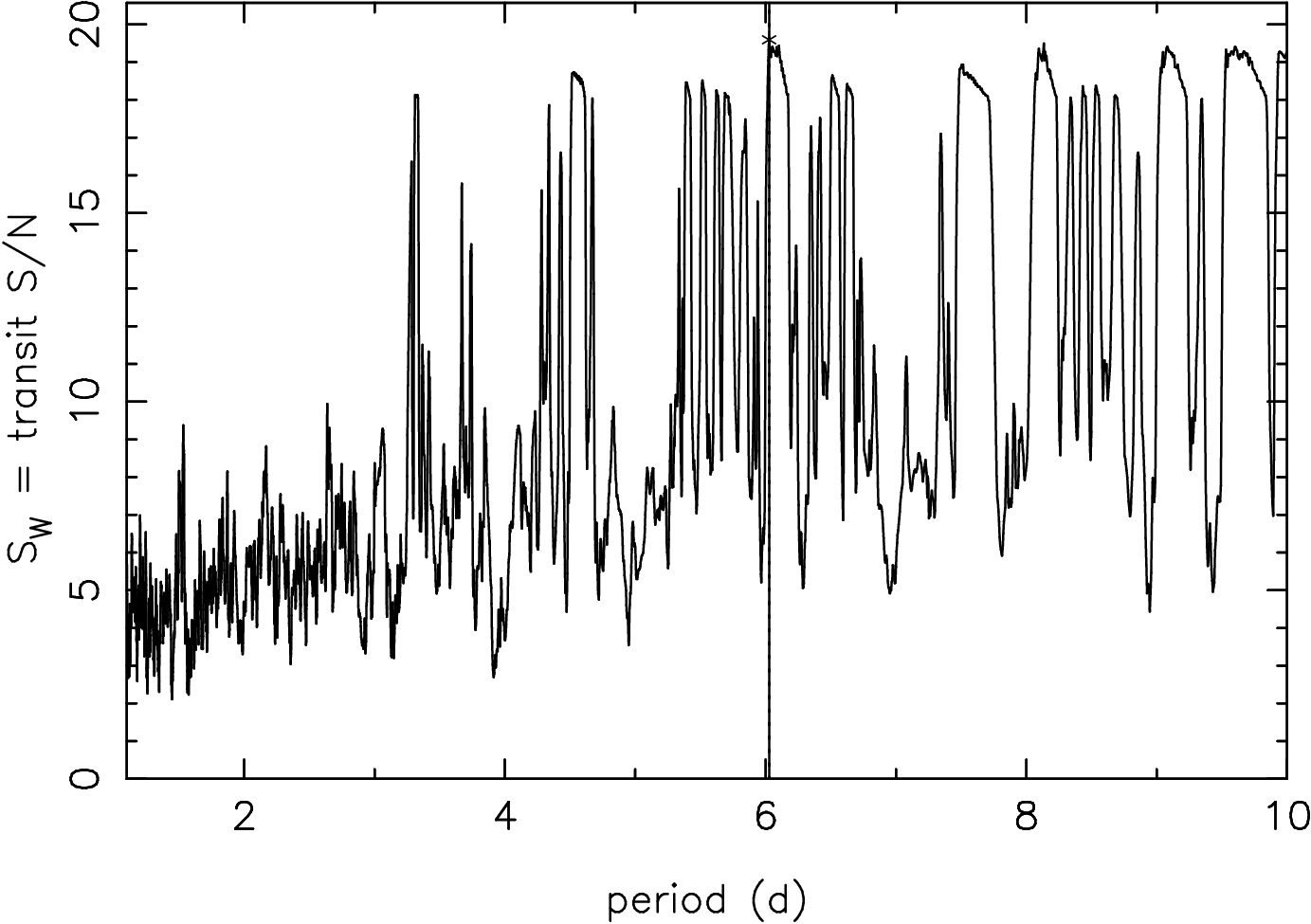} \\
      \\
      \includegraphics[angle=270,width=5.2cm]{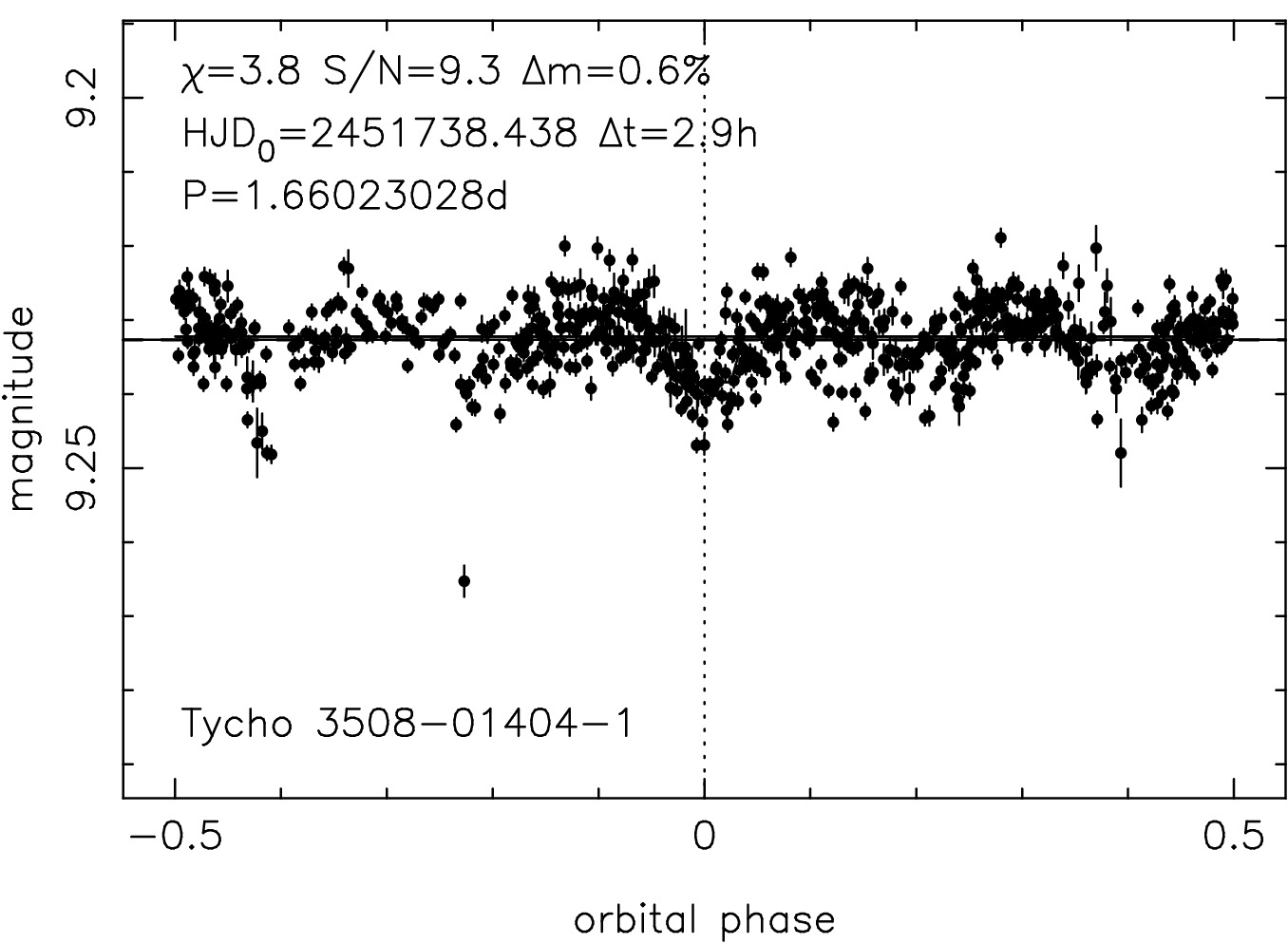} &
      \includegraphics[angle=270,width=5.2cm]{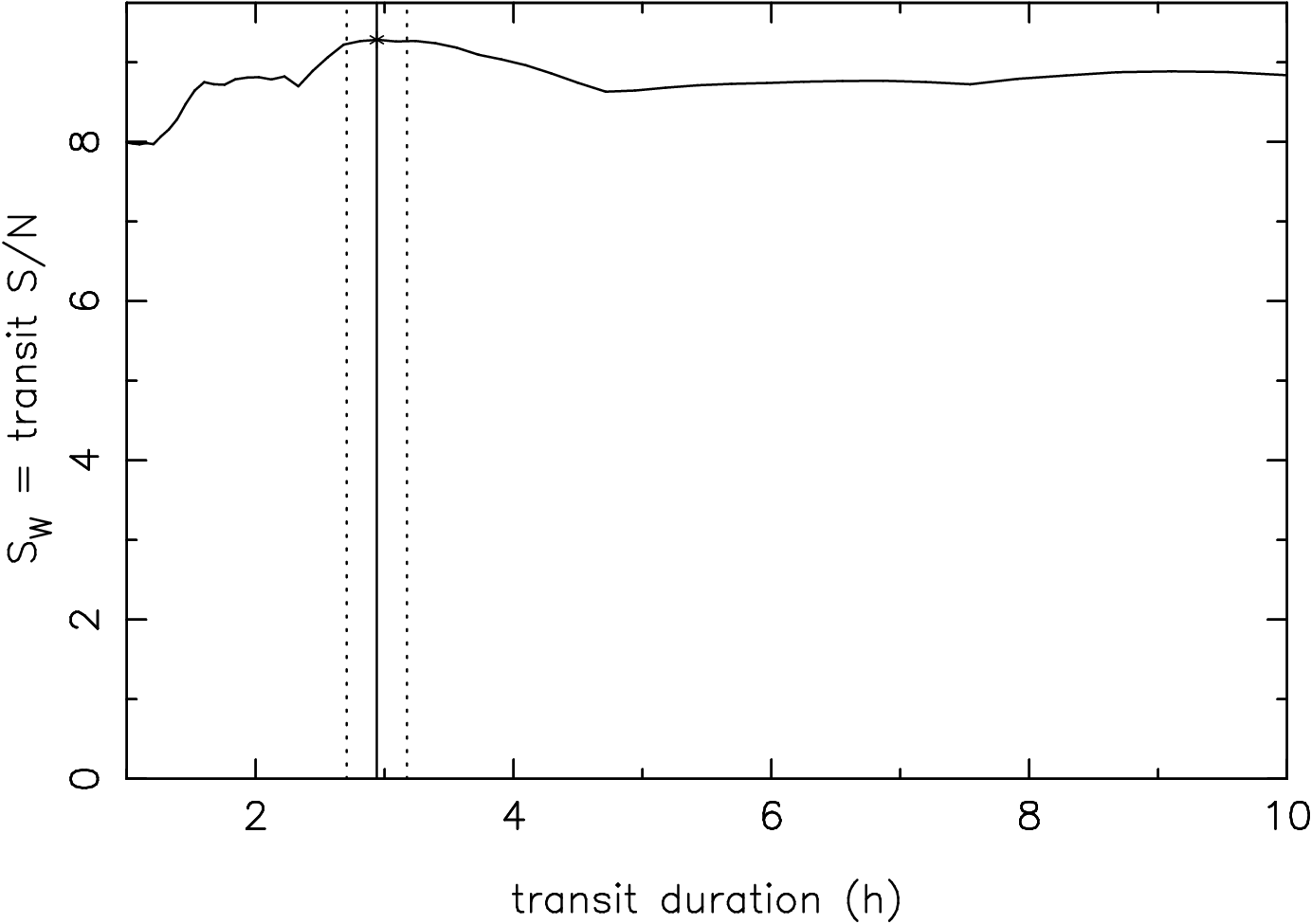} &
      \includegraphics[angle=270,width=5.2cm]{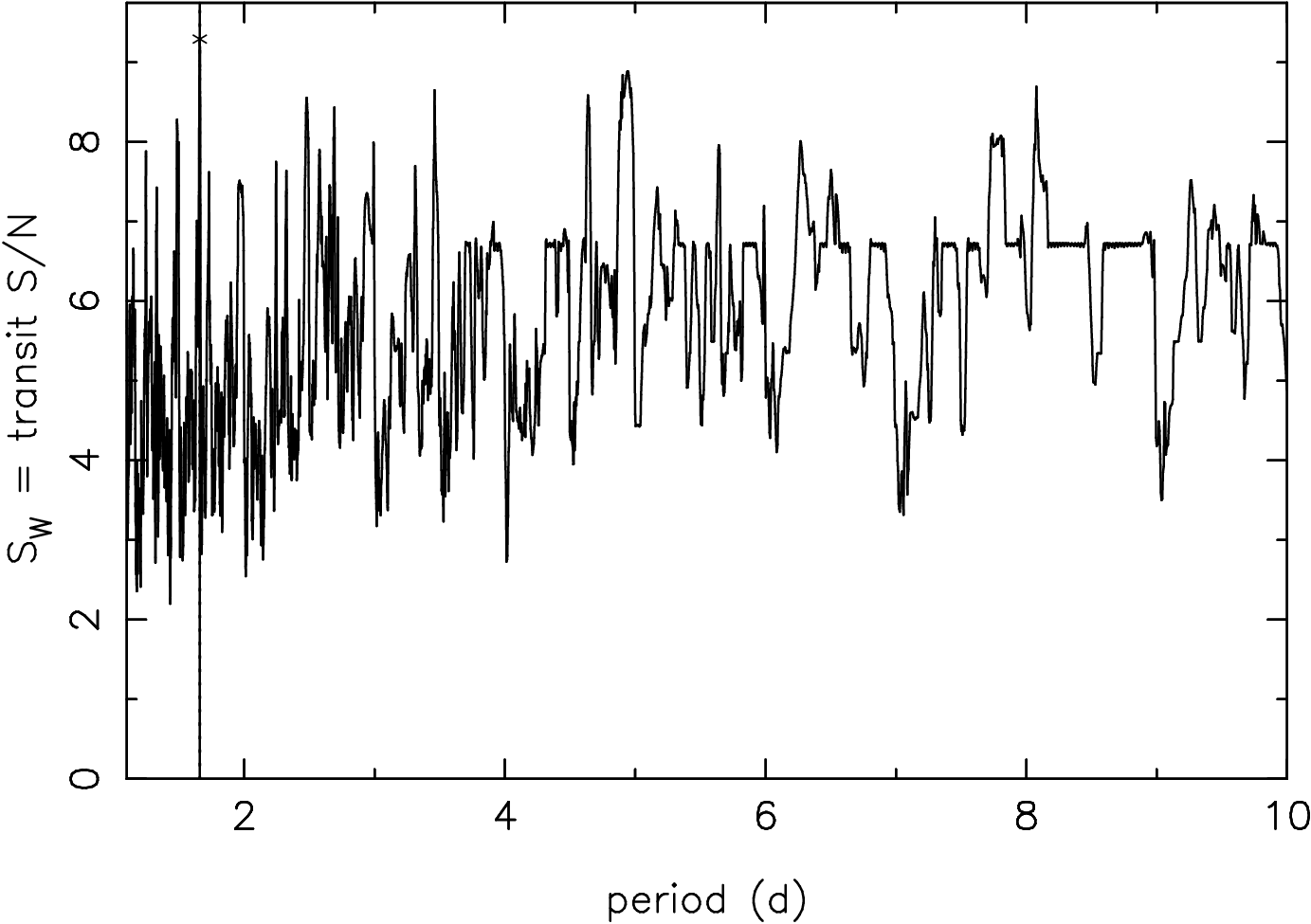} \\
      \\
      \includegraphics[angle=270,width=5.2cm]{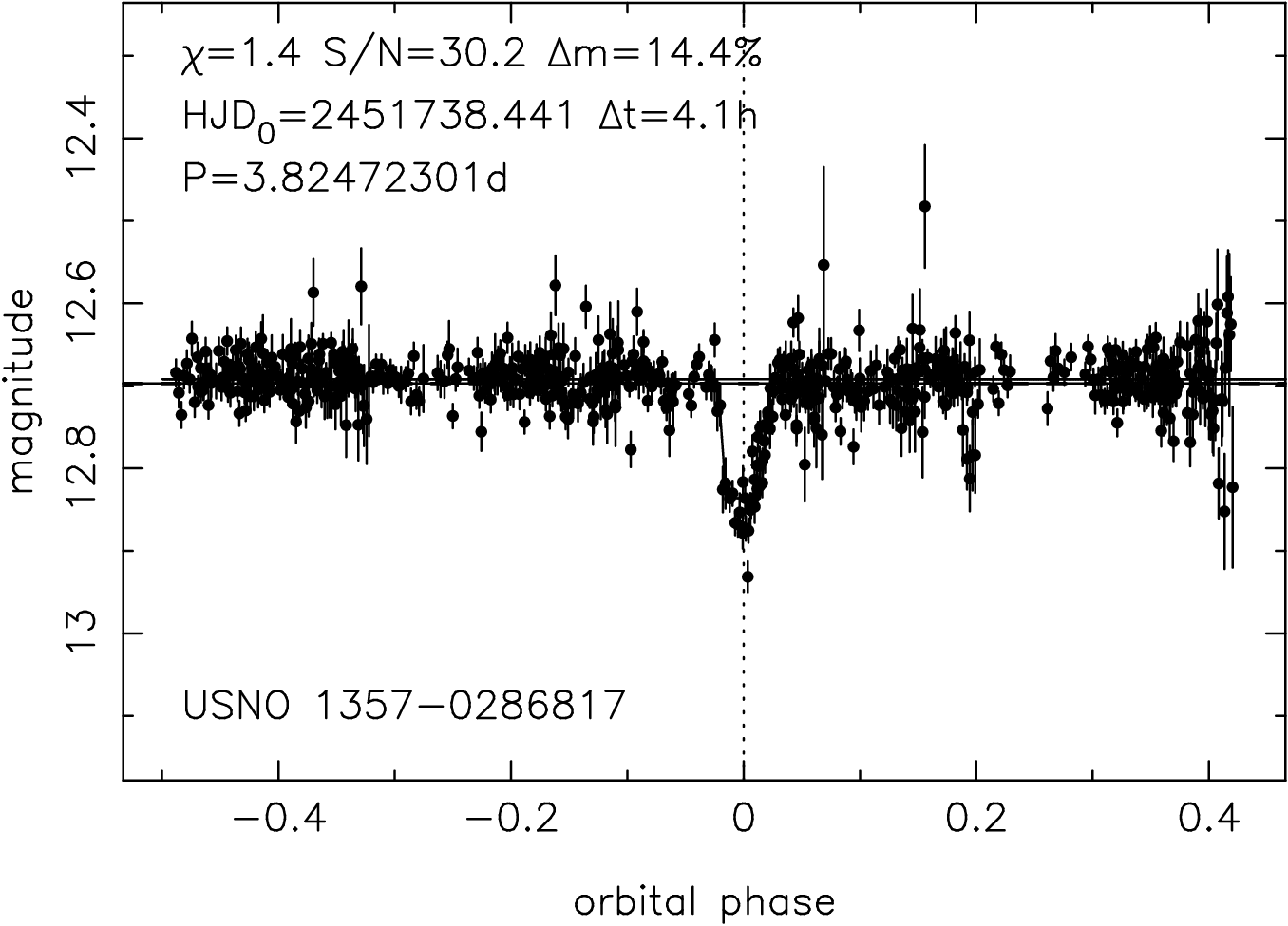} &
      \includegraphics[angle=270,width=5.2cm]{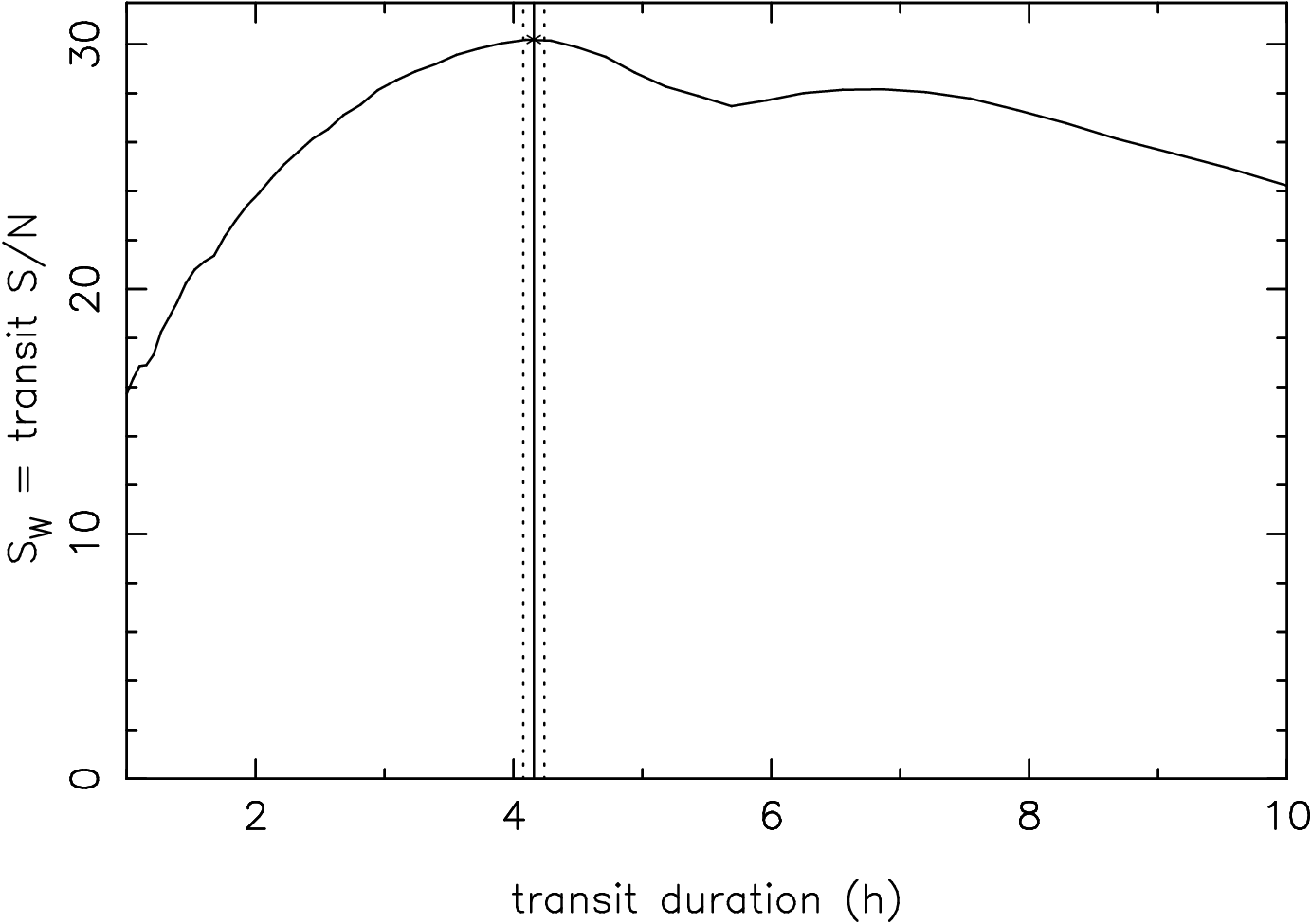} &
      \includegraphics[angle=270,width=5.2cm]{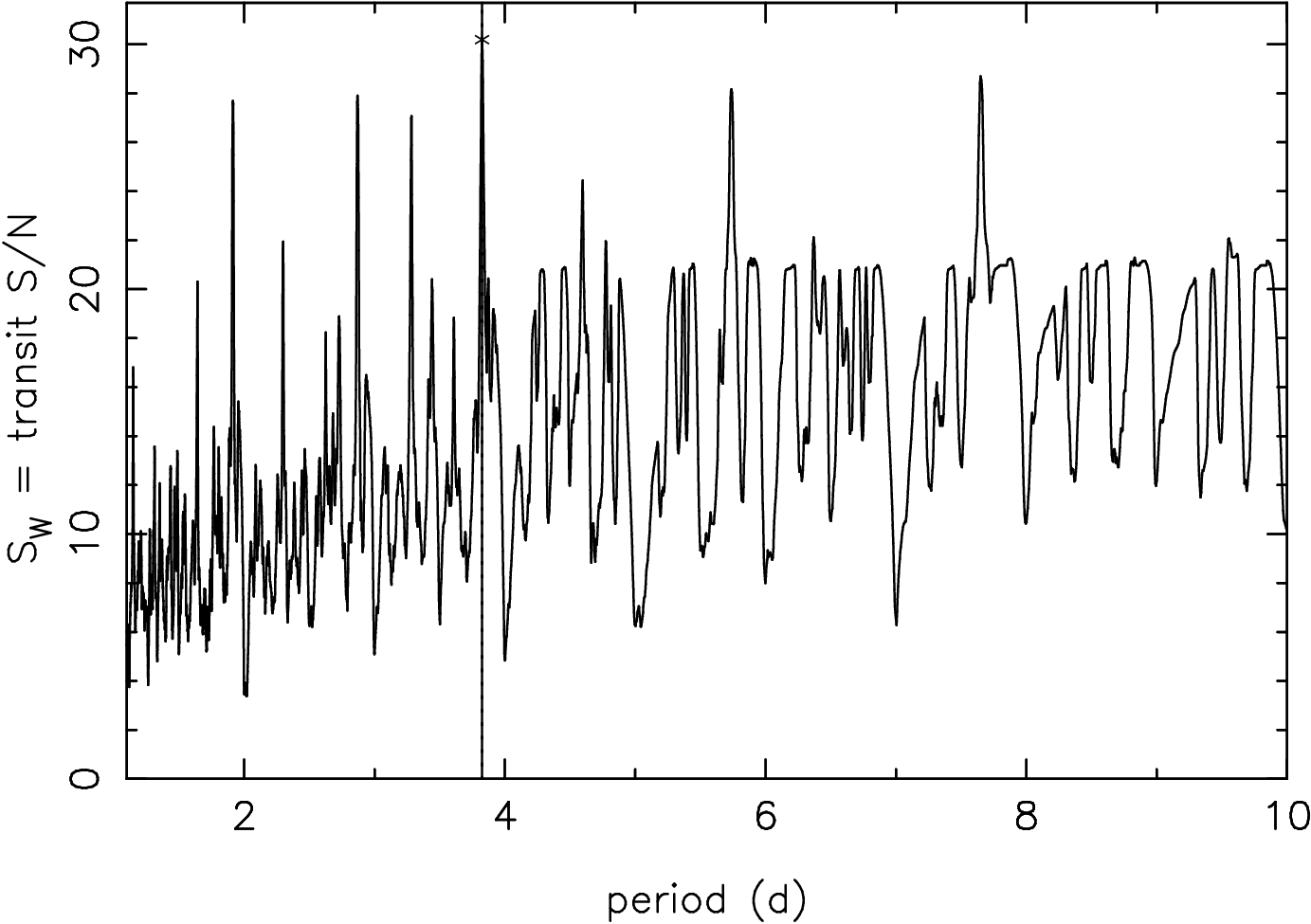} \\
      \\
      \includegraphics[angle=270,width=5.2cm]{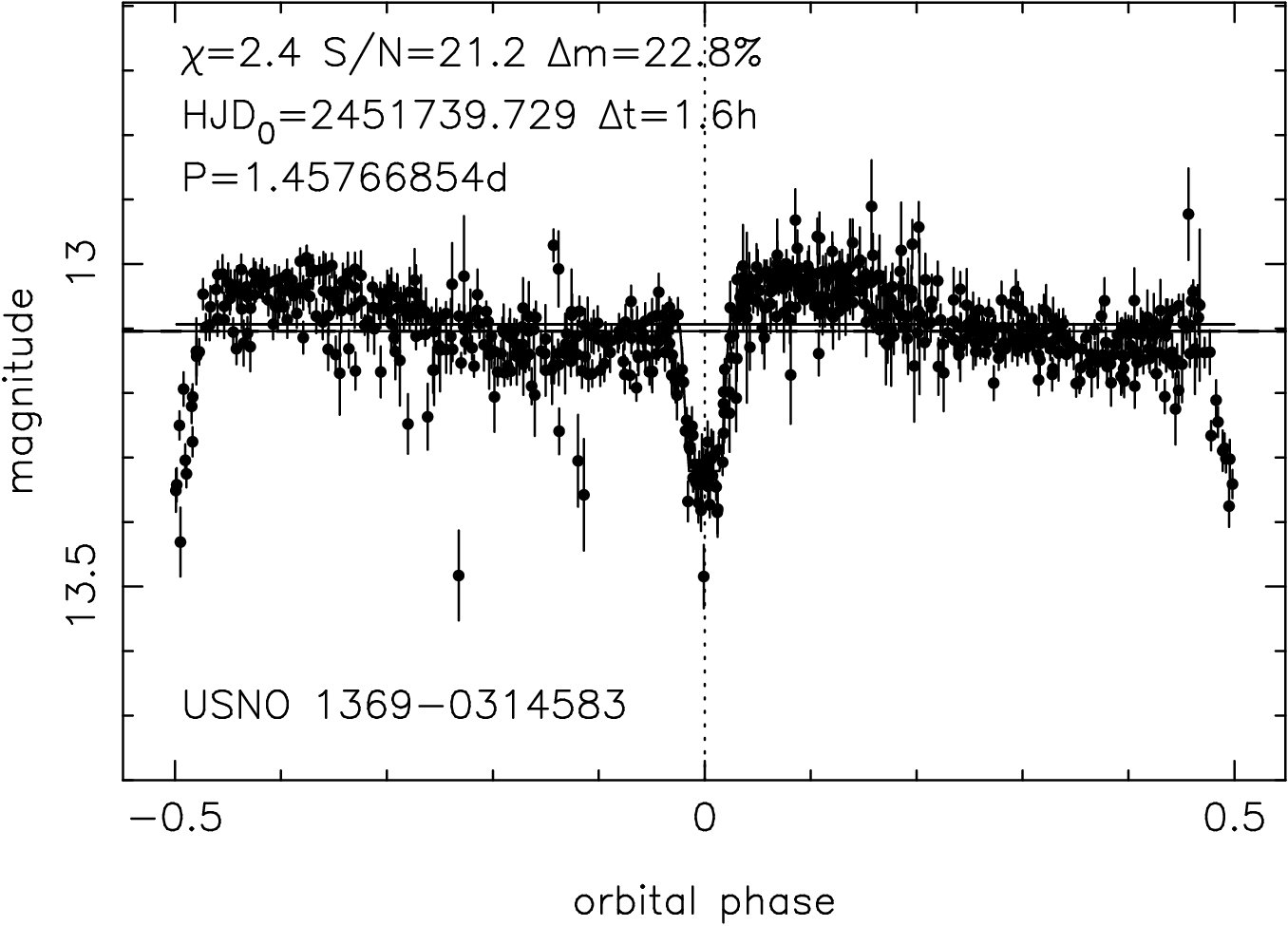} &
      \includegraphics[angle=270,width=5.2cm]{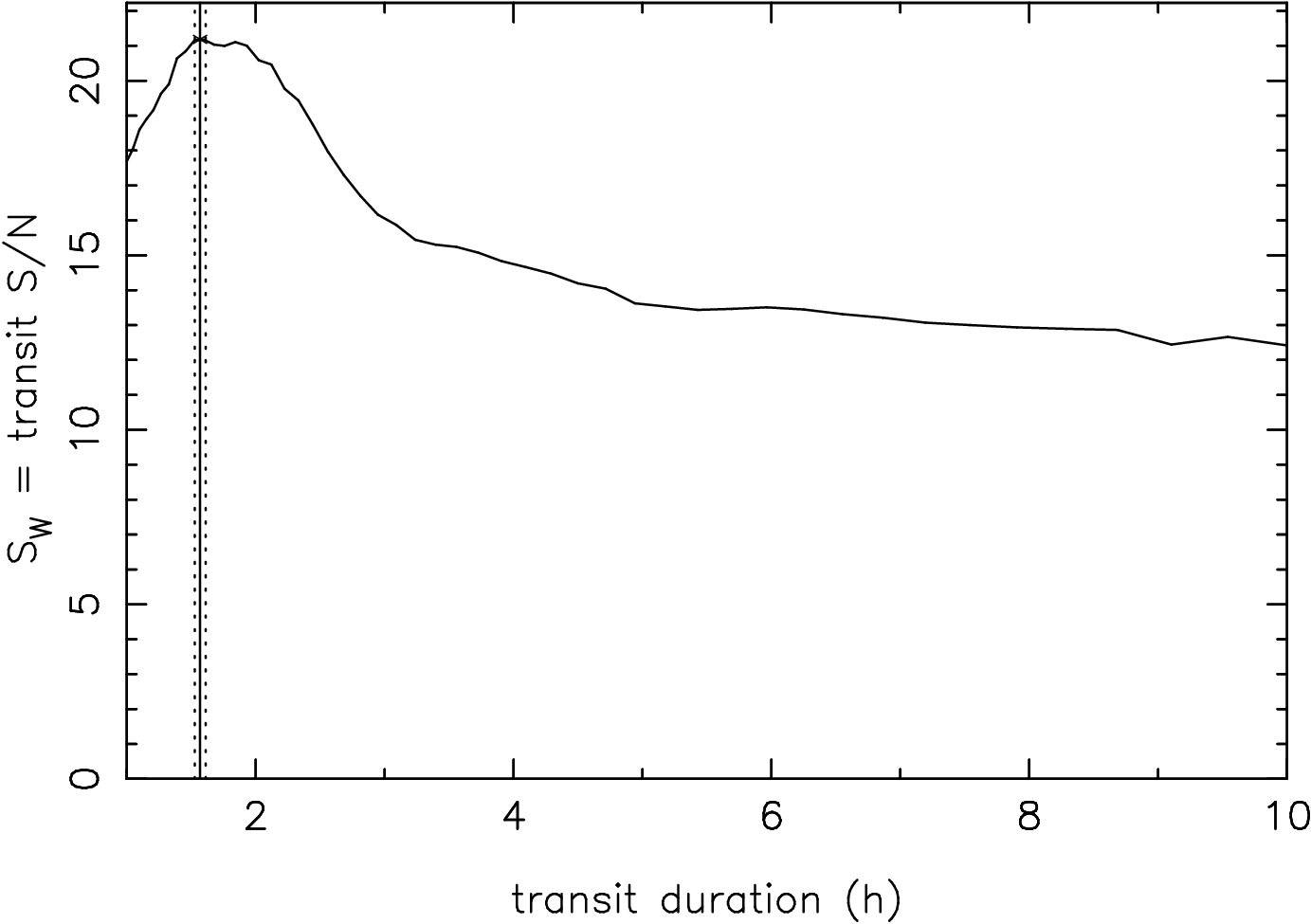} &
      \includegraphics[angle=270,width=5.2cm]{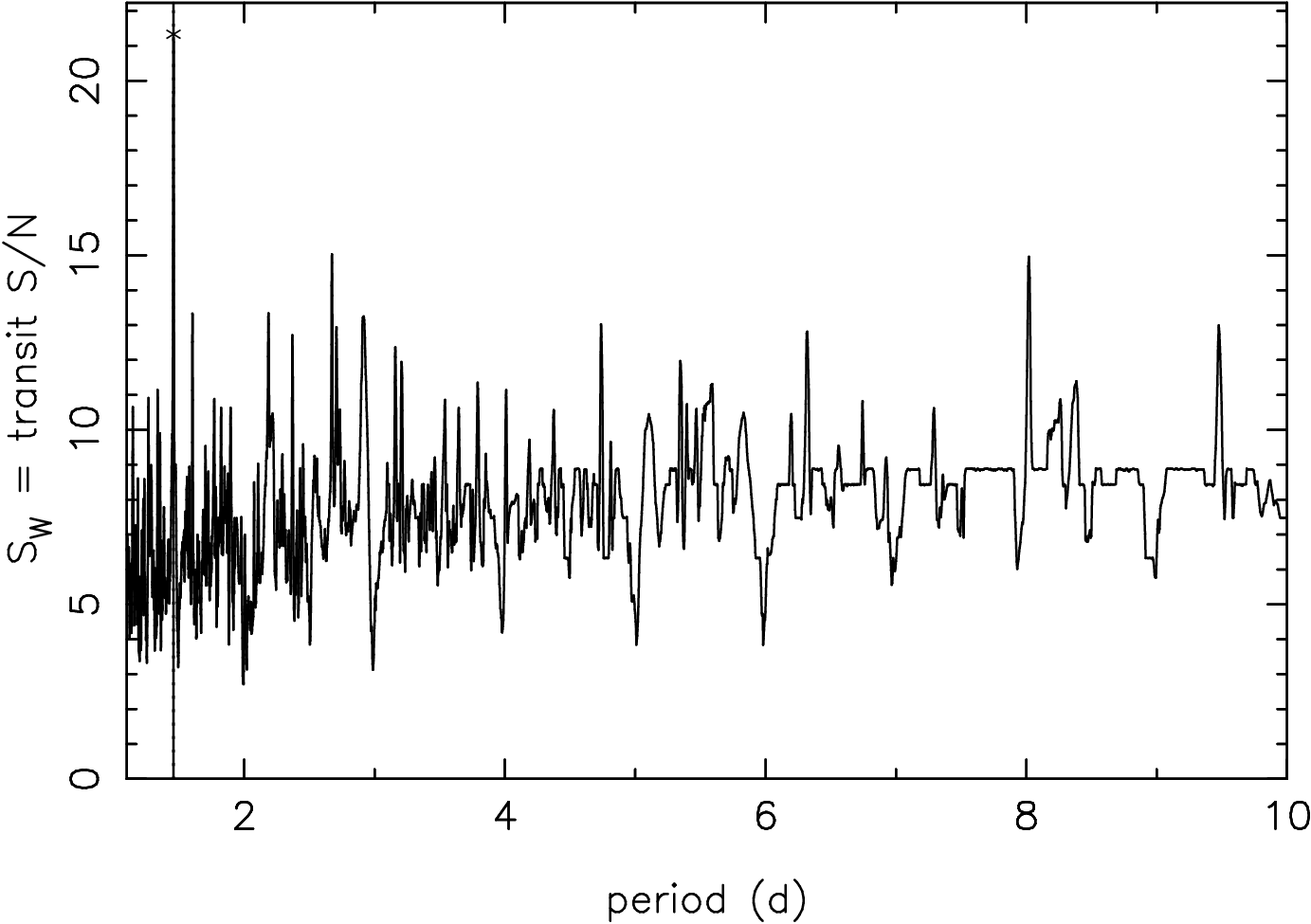} \\
      \\
      \includegraphics[angle=270,width=5.2cm]{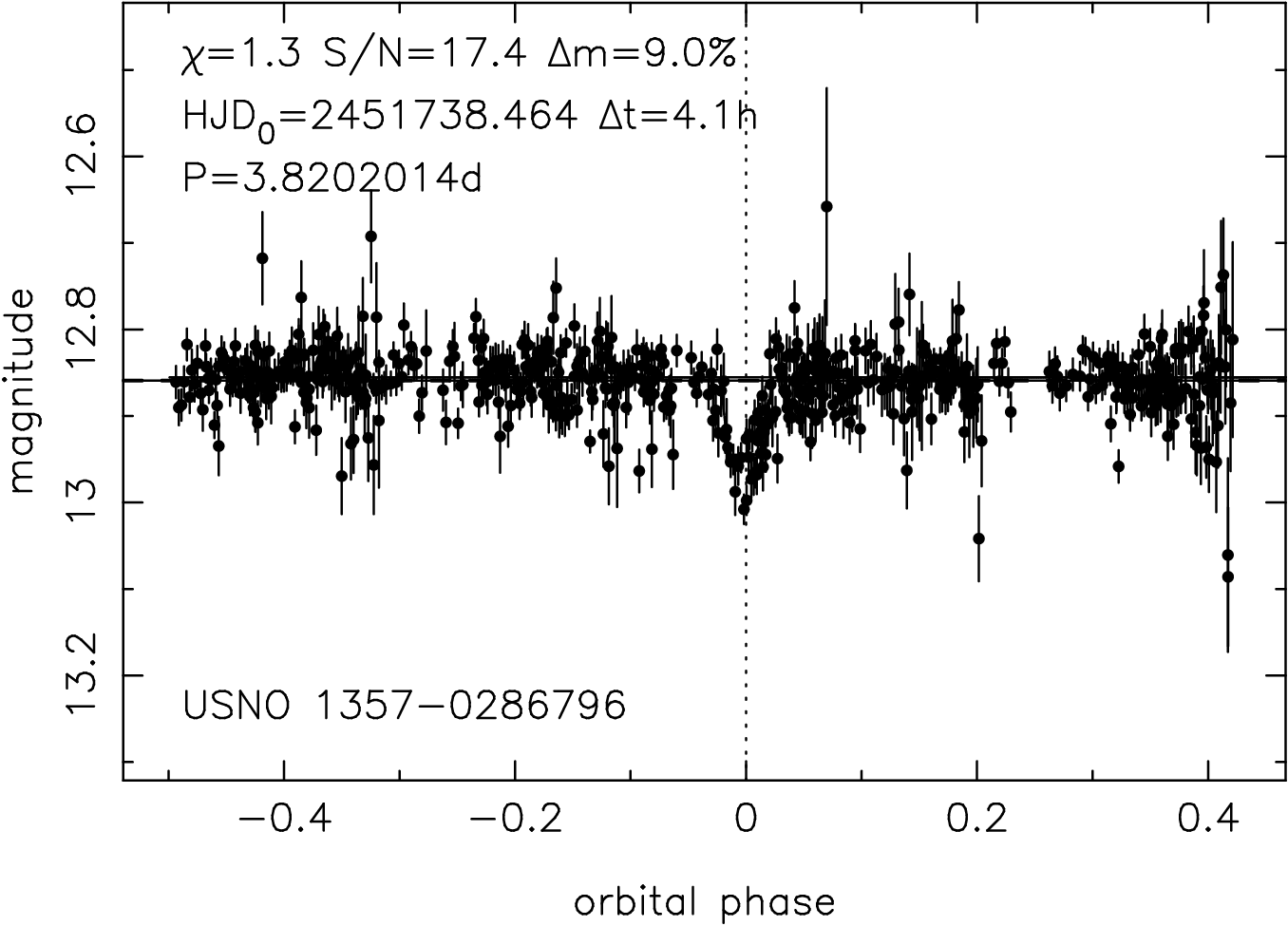} &
      \includegraphics[angle=270,width=5.2cm]{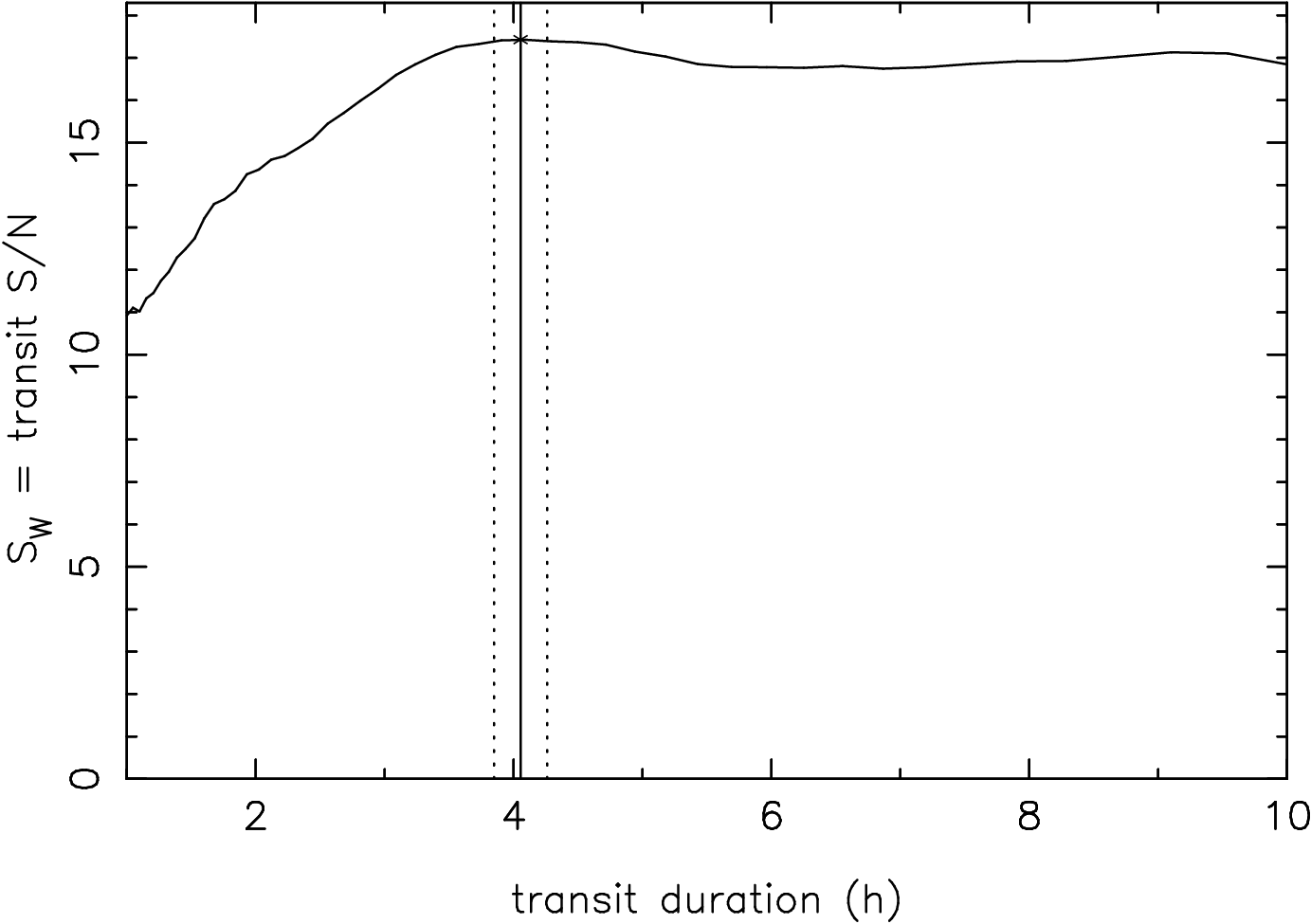} &
      \includegraphics[angle=270,width=5.2cm]{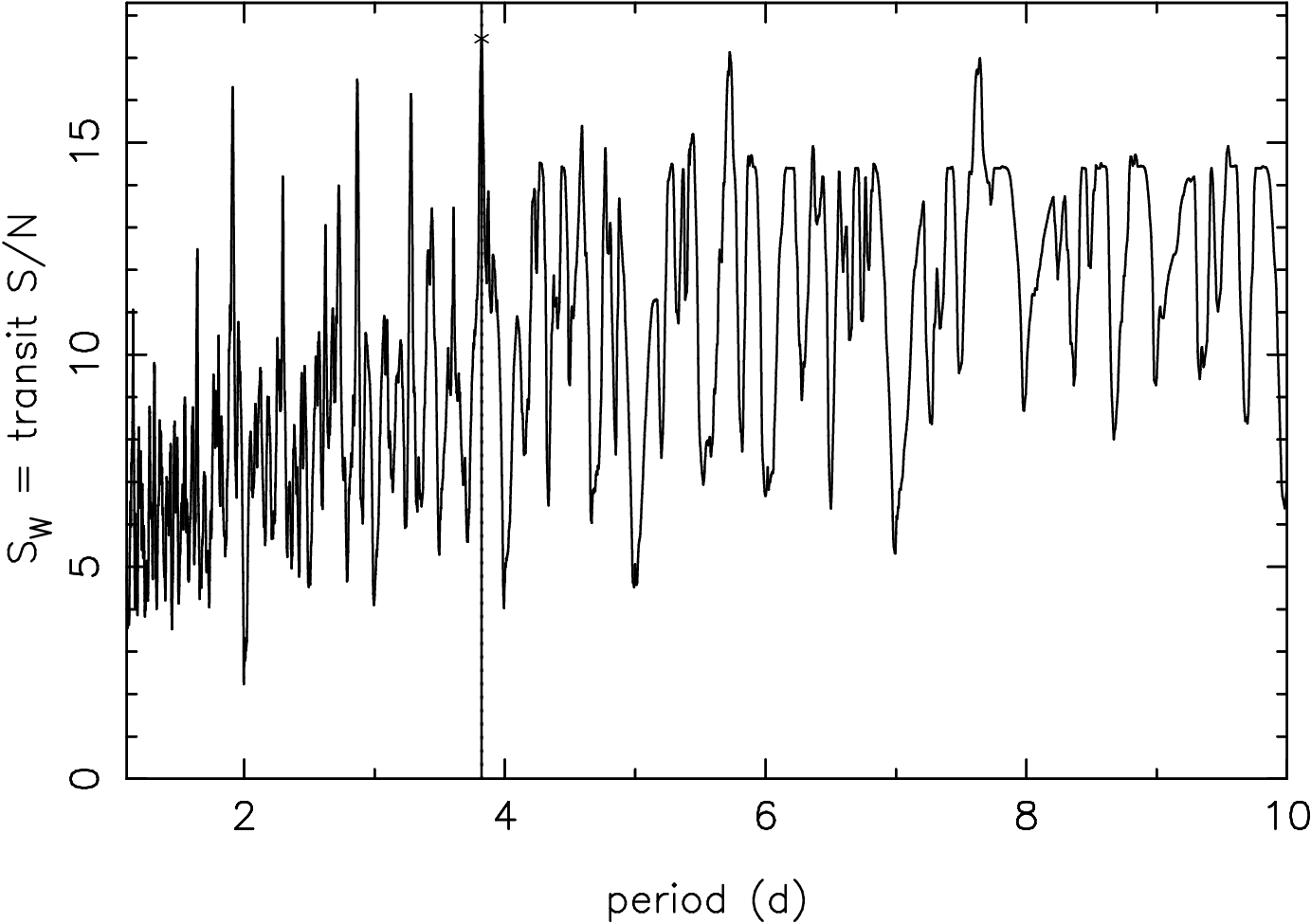} \\
    \end{tabular}
  \end{center}
  \caption{Transit candidates 6--10 (top to bottom) from the Draco
field.}
\end{figure*}

\begin{figure*}
  \begin{center}
    \begin{tabular}{ccc}
      \includegraphics[angle=270,width=5.2cm]{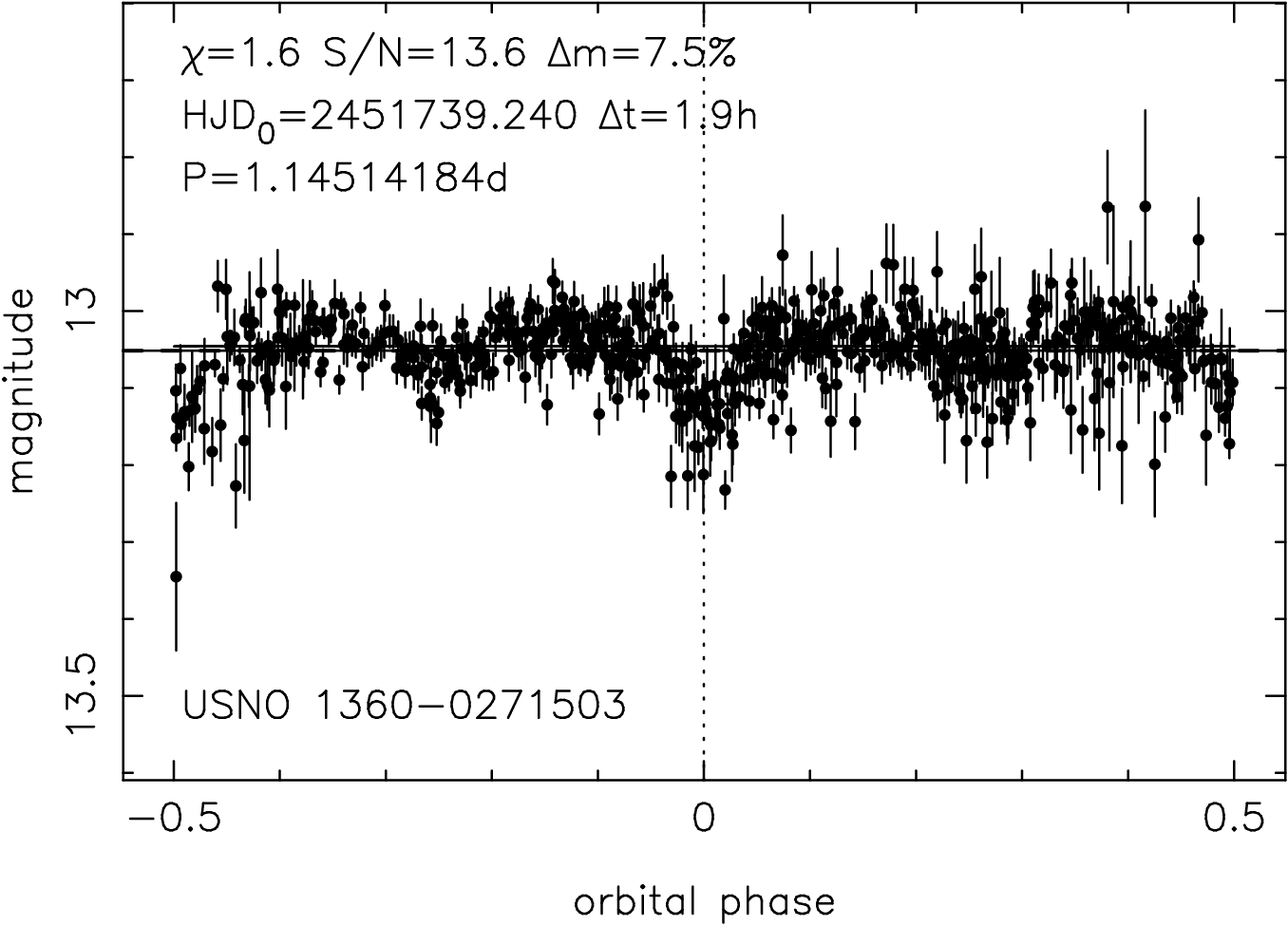} &
      \includegraphics[angle=270,width=5.2cm]{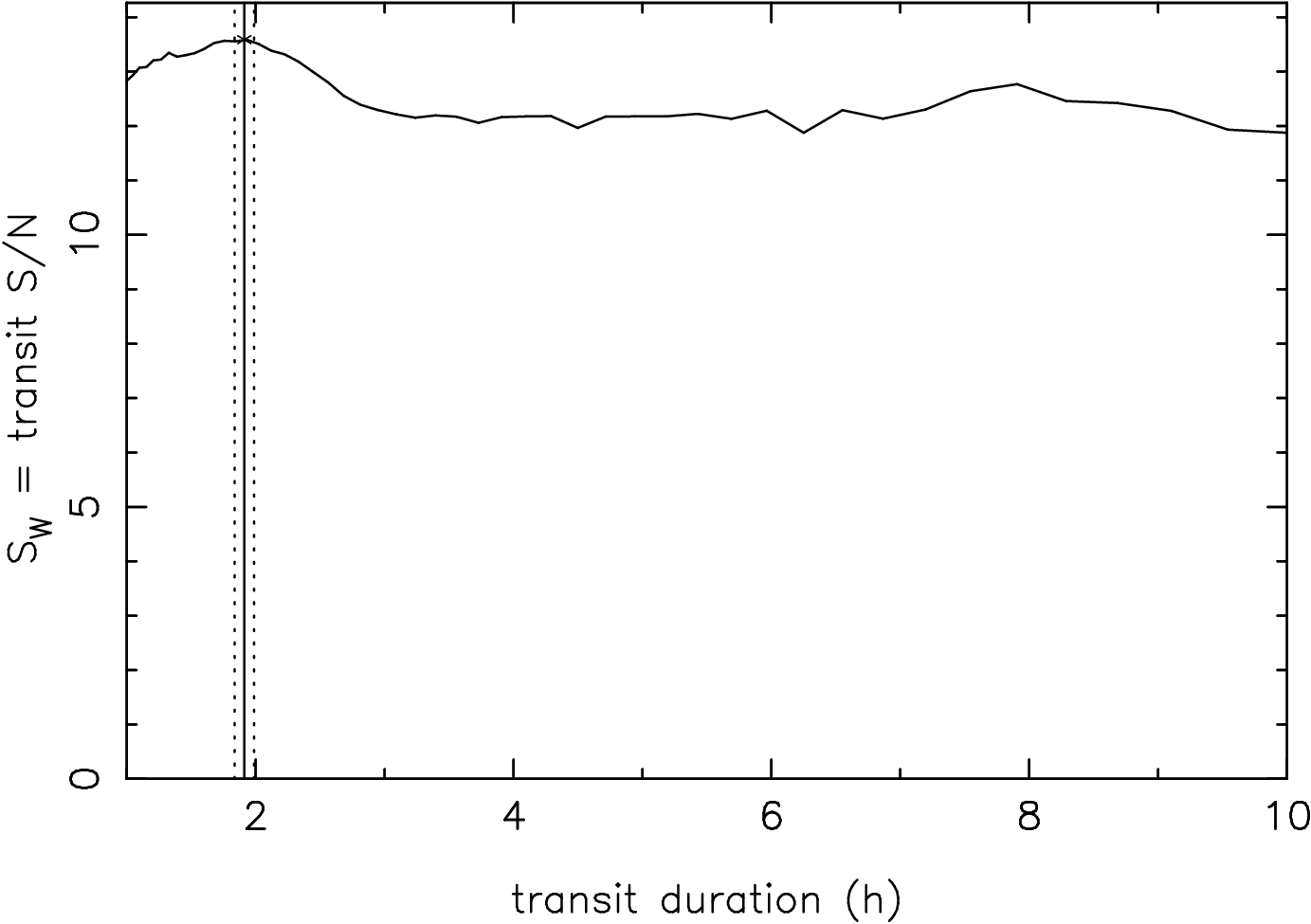} &
      \includegraphics[angle=270,width=5.2cm]{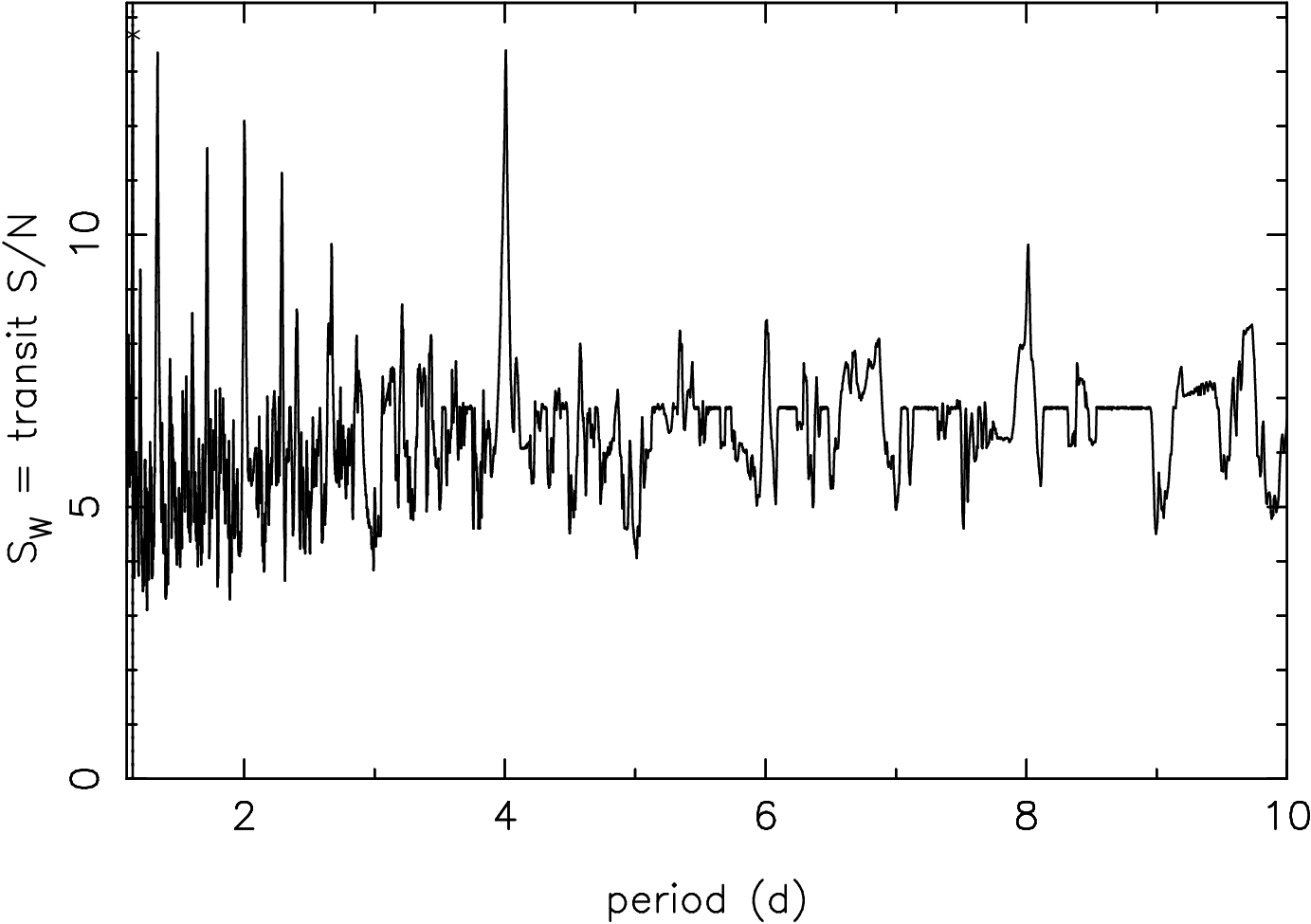} \\
      \\
      \includegraphics[angle=270,width=5.2cm]{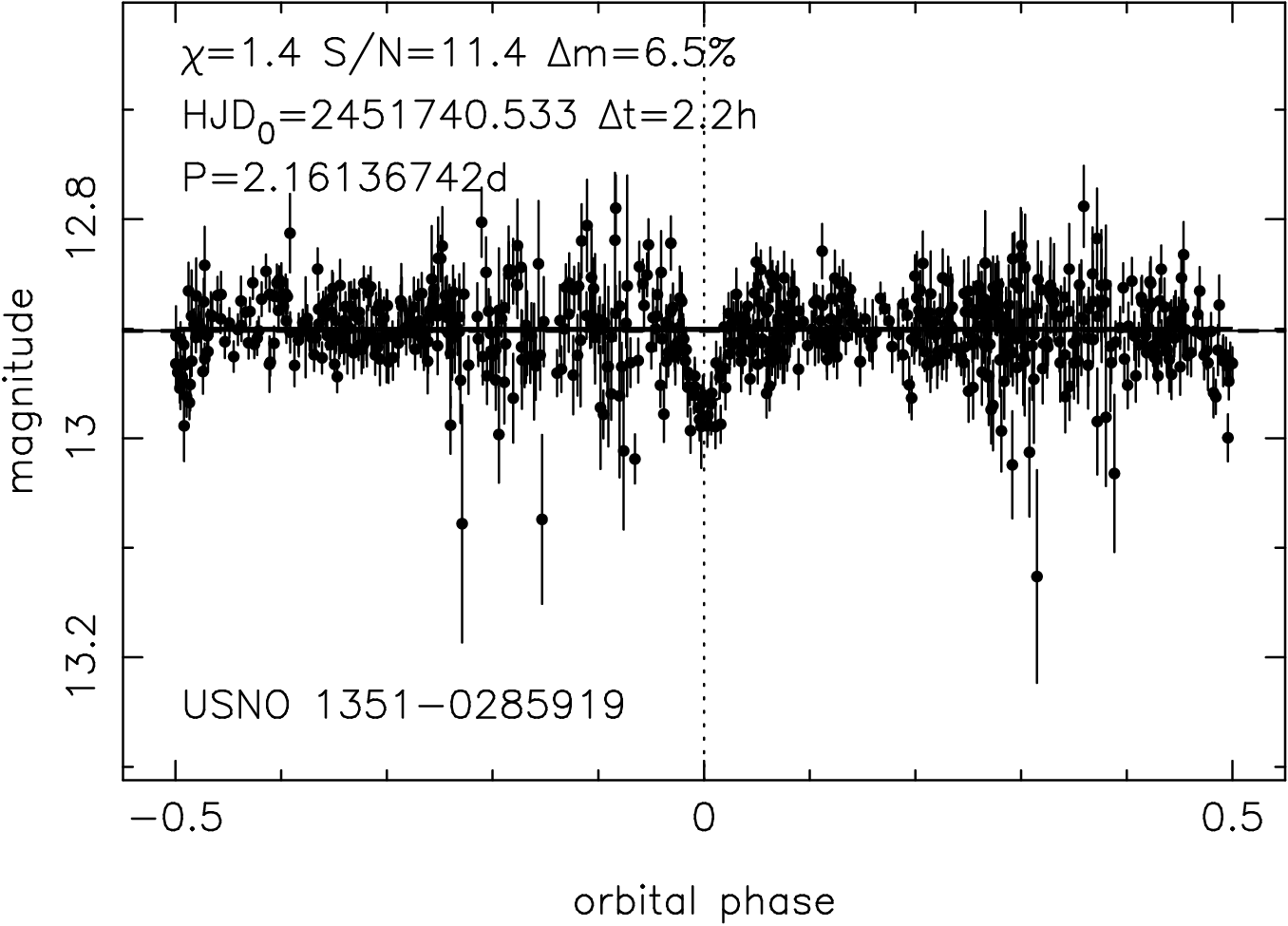} &
      \includegraphics[angle=270,width=5.2cm]{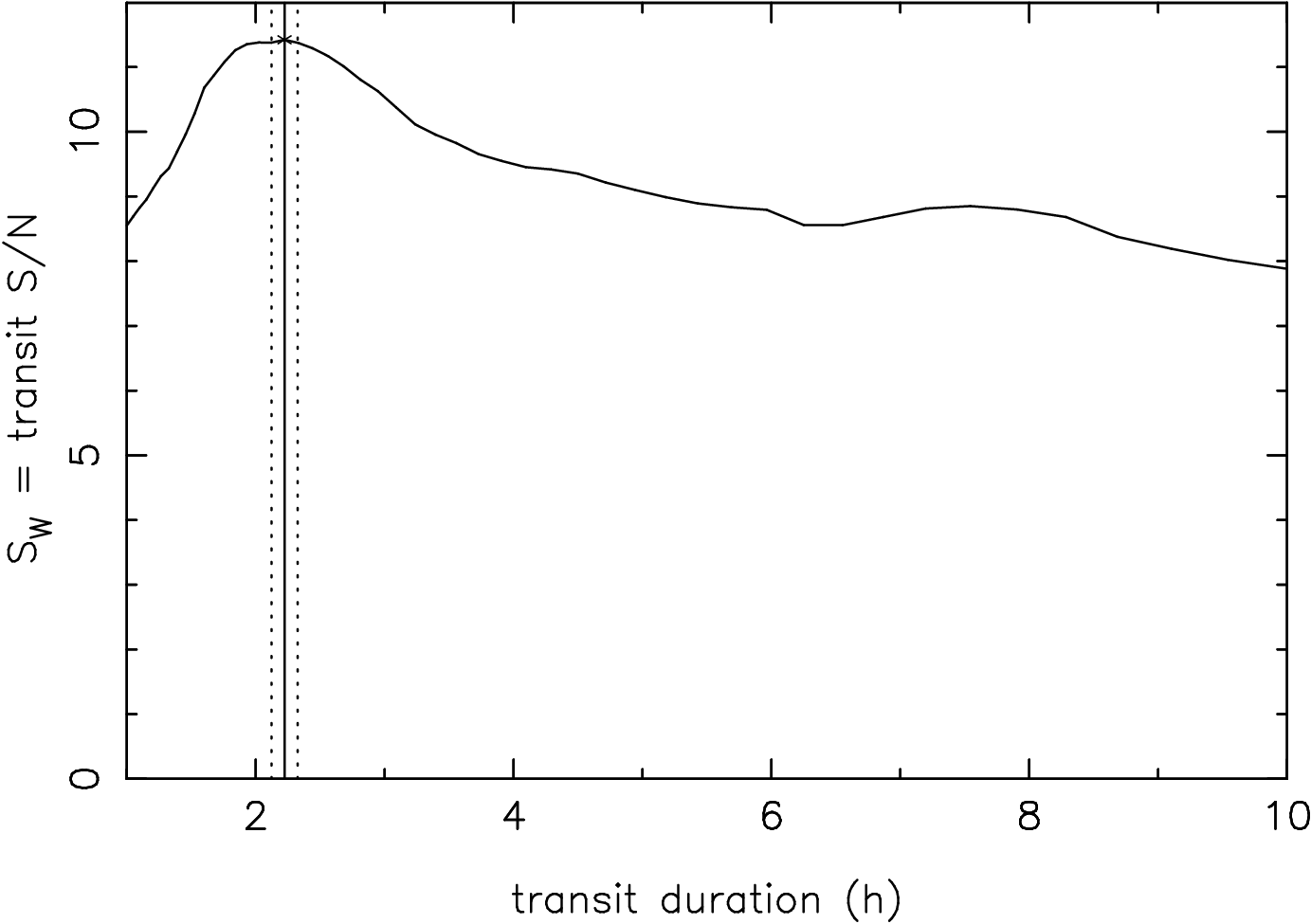} &
      \includegraphics[angle=270,width=5.2cm]{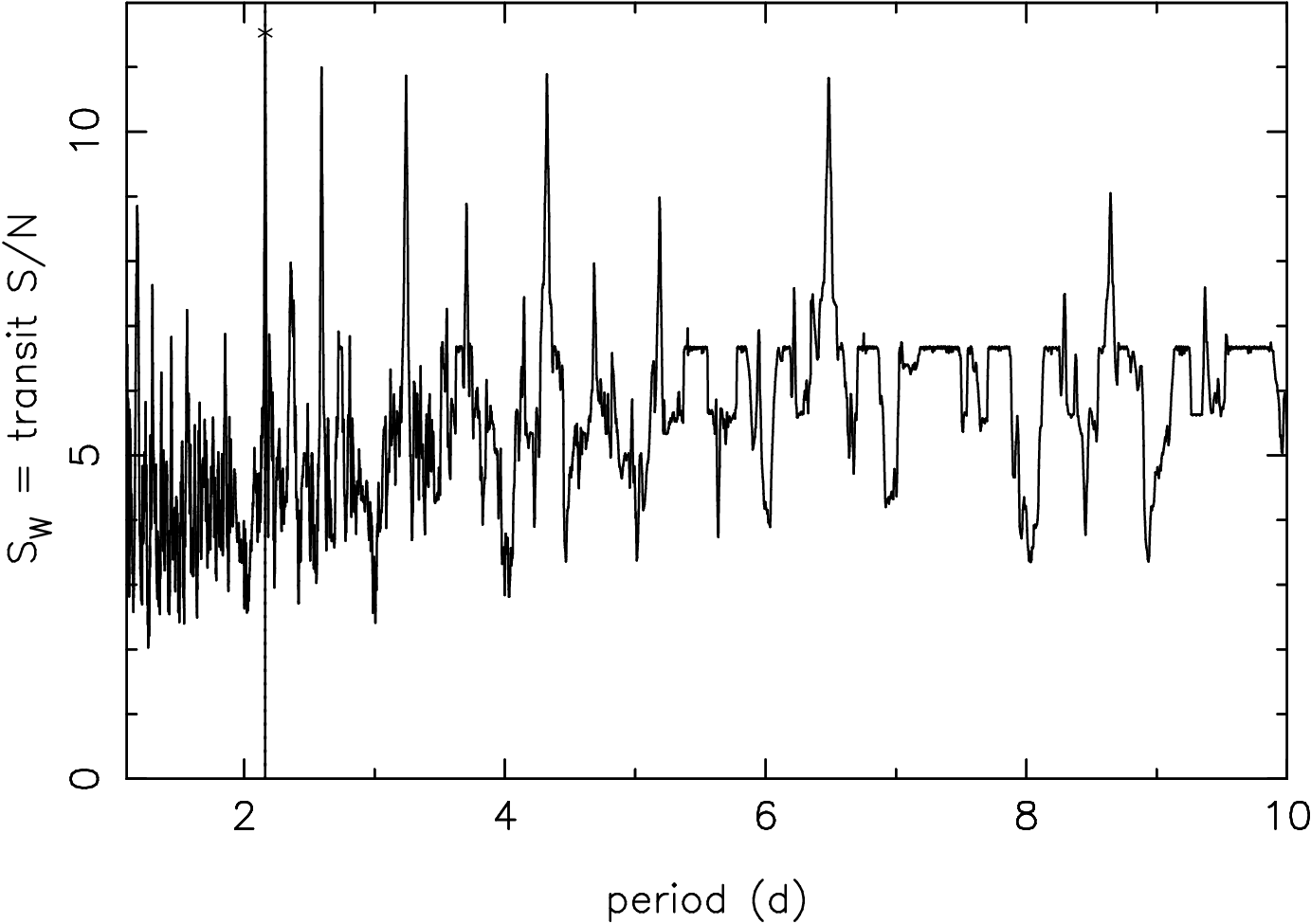} \\
      \\
      \includegraphics[angle=270,width=5.2cm]{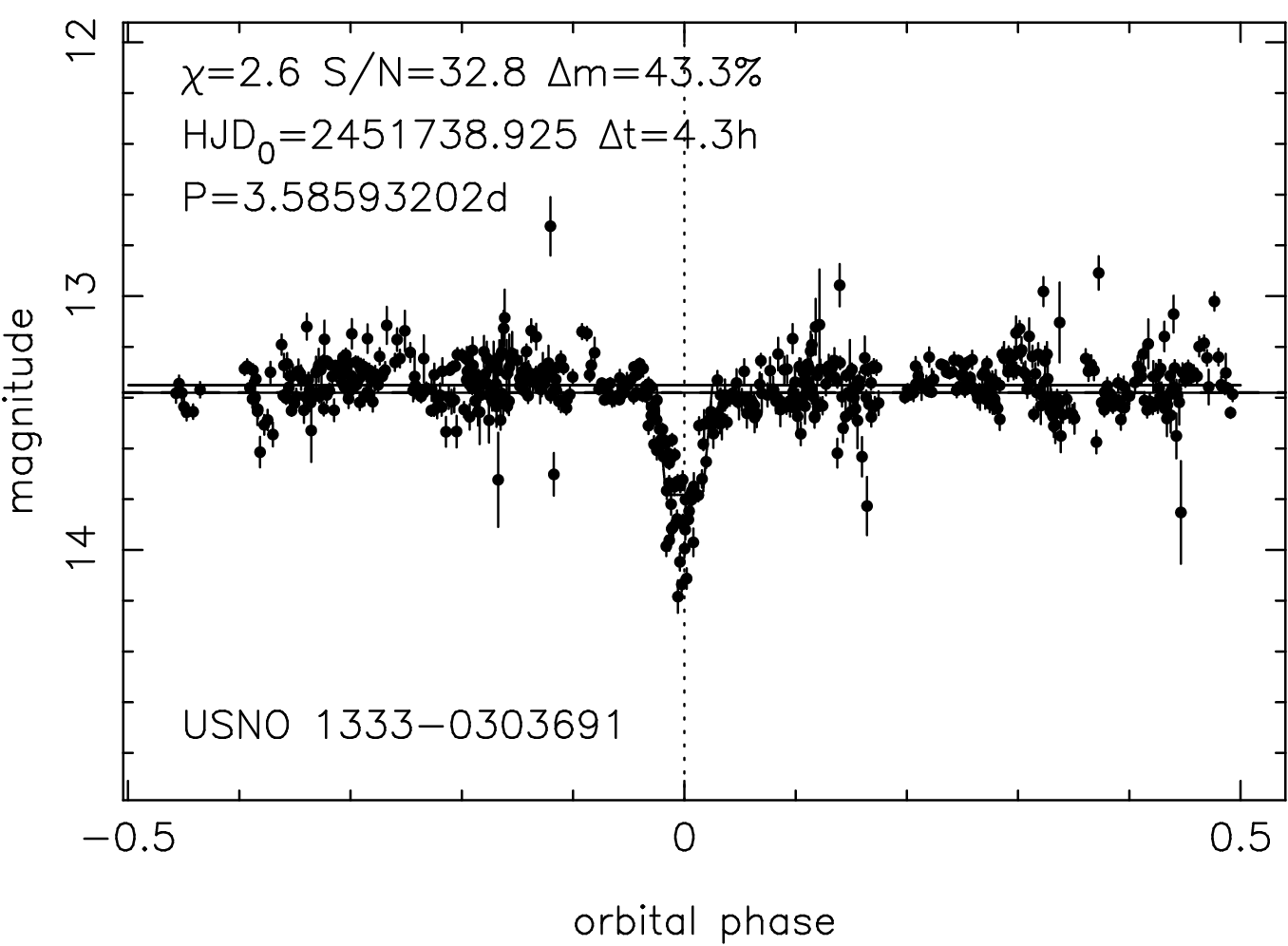} &
      \includegraphics[angle=270,width=5.2cm]{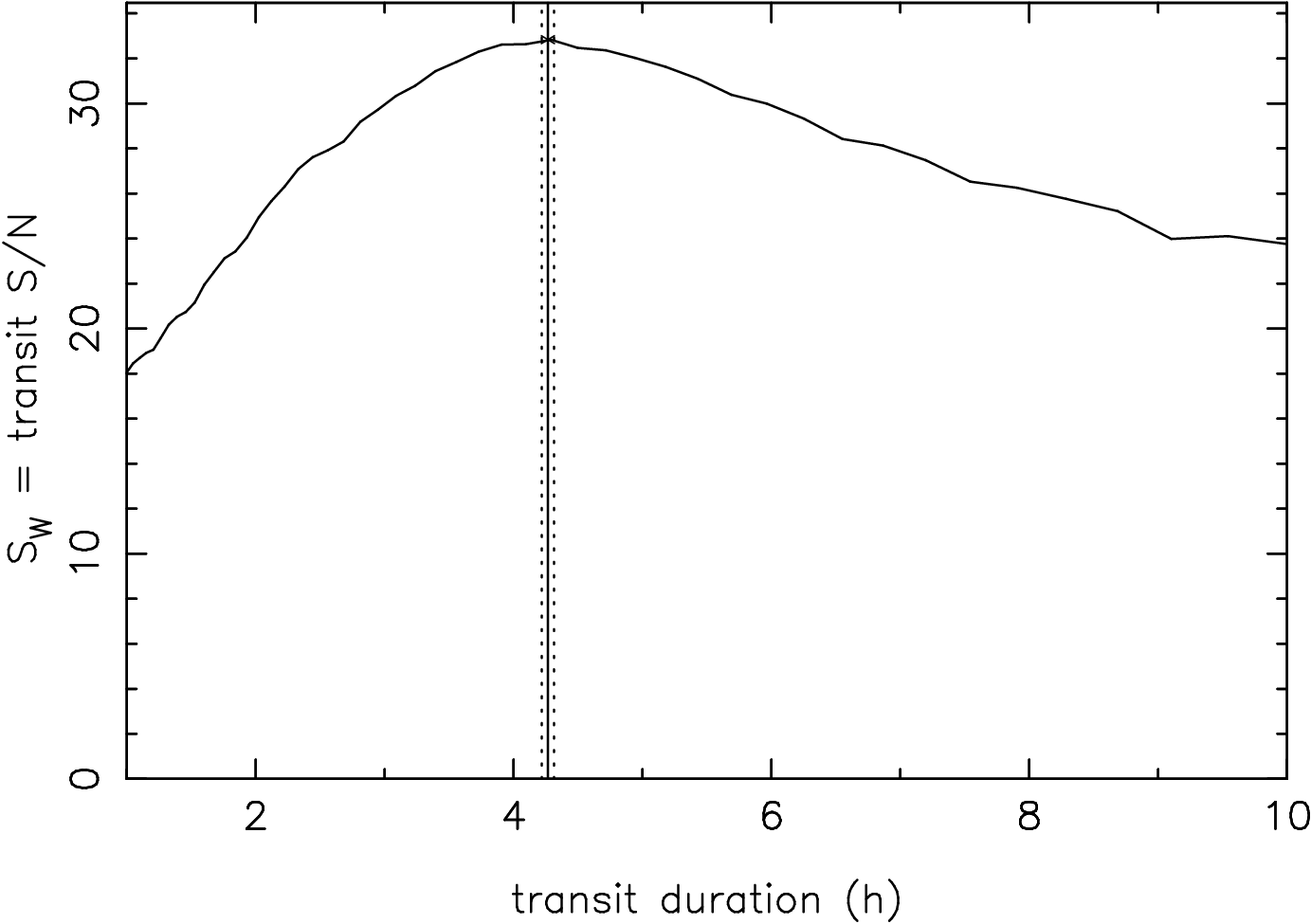} &
      \includegraphics[angle=270,width=5.2cm]{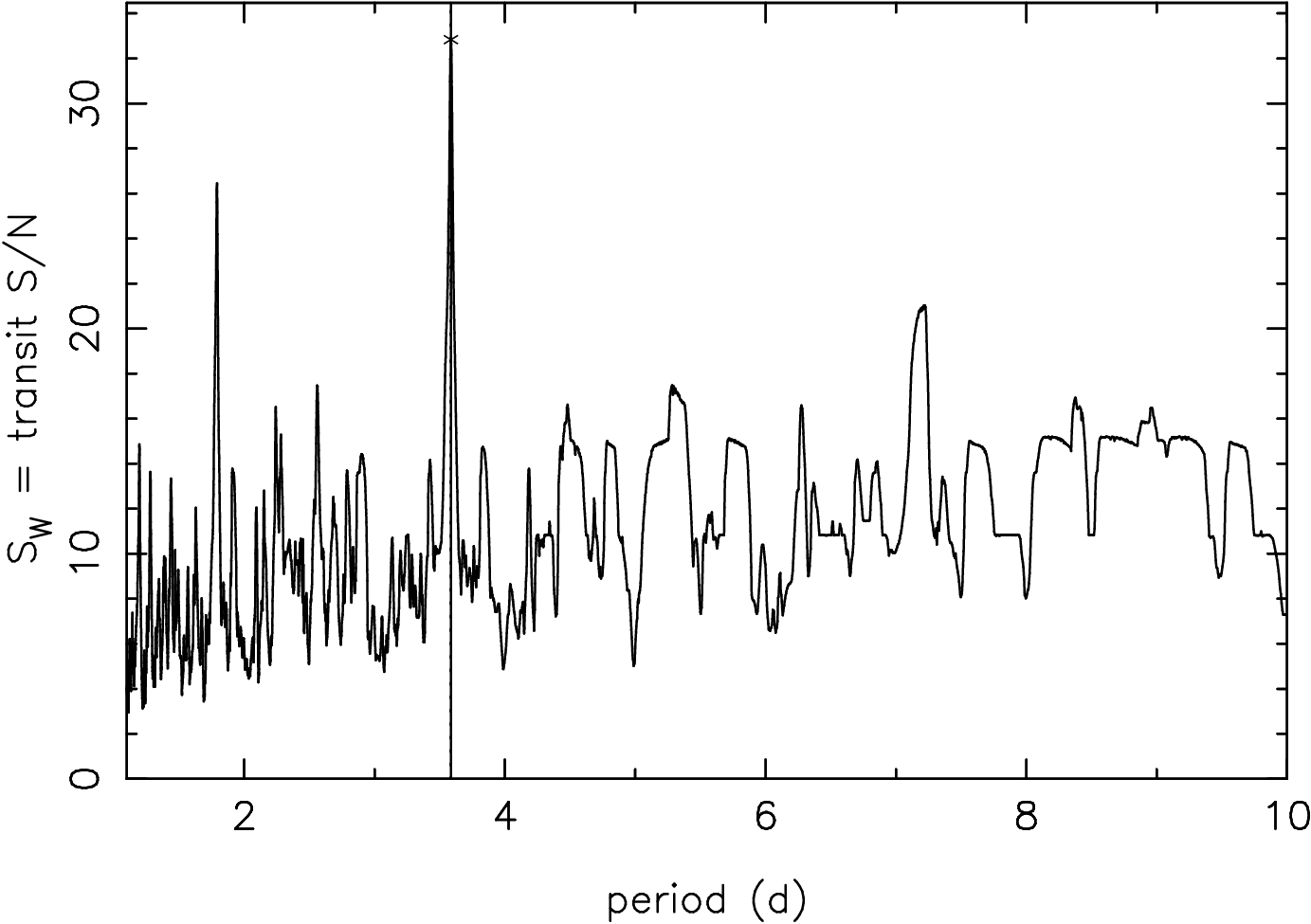} \\
      \\
      \includegraphics[angle=270,width=5.2cm]{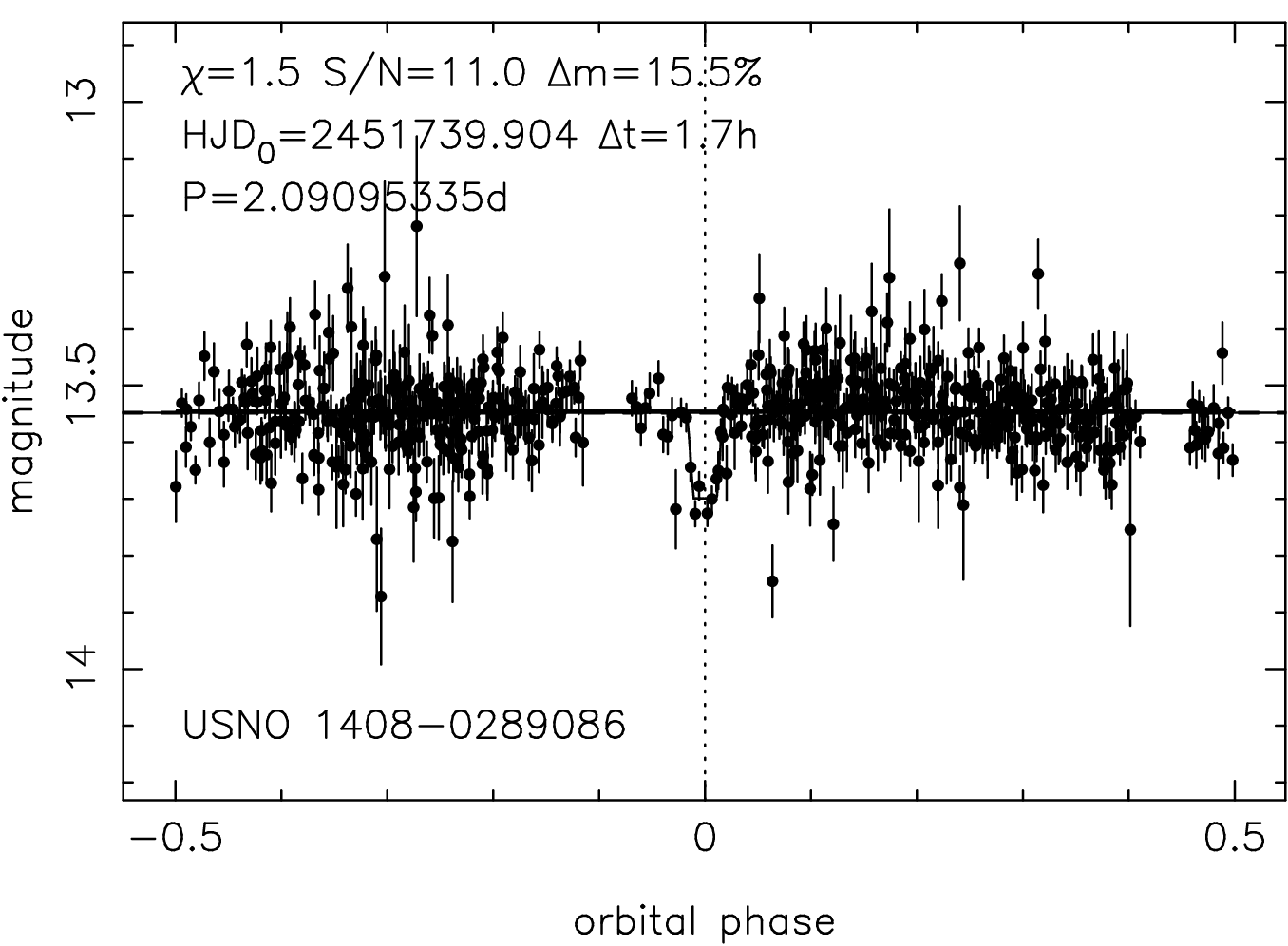} &
      \includegraphics[angle=270,width=5.2cm]{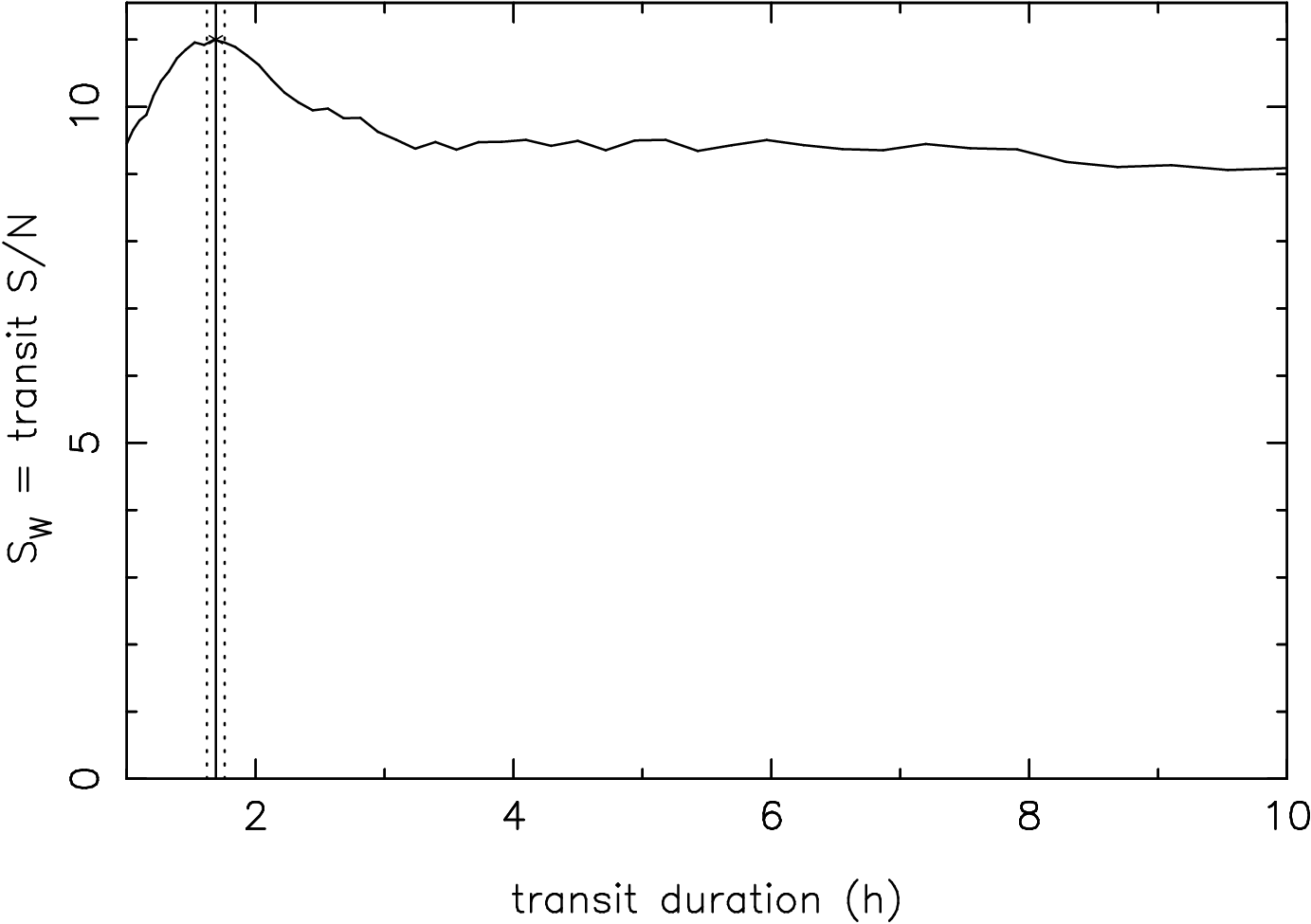} &
      \includegraphics[angle=270,width=5.2cm]{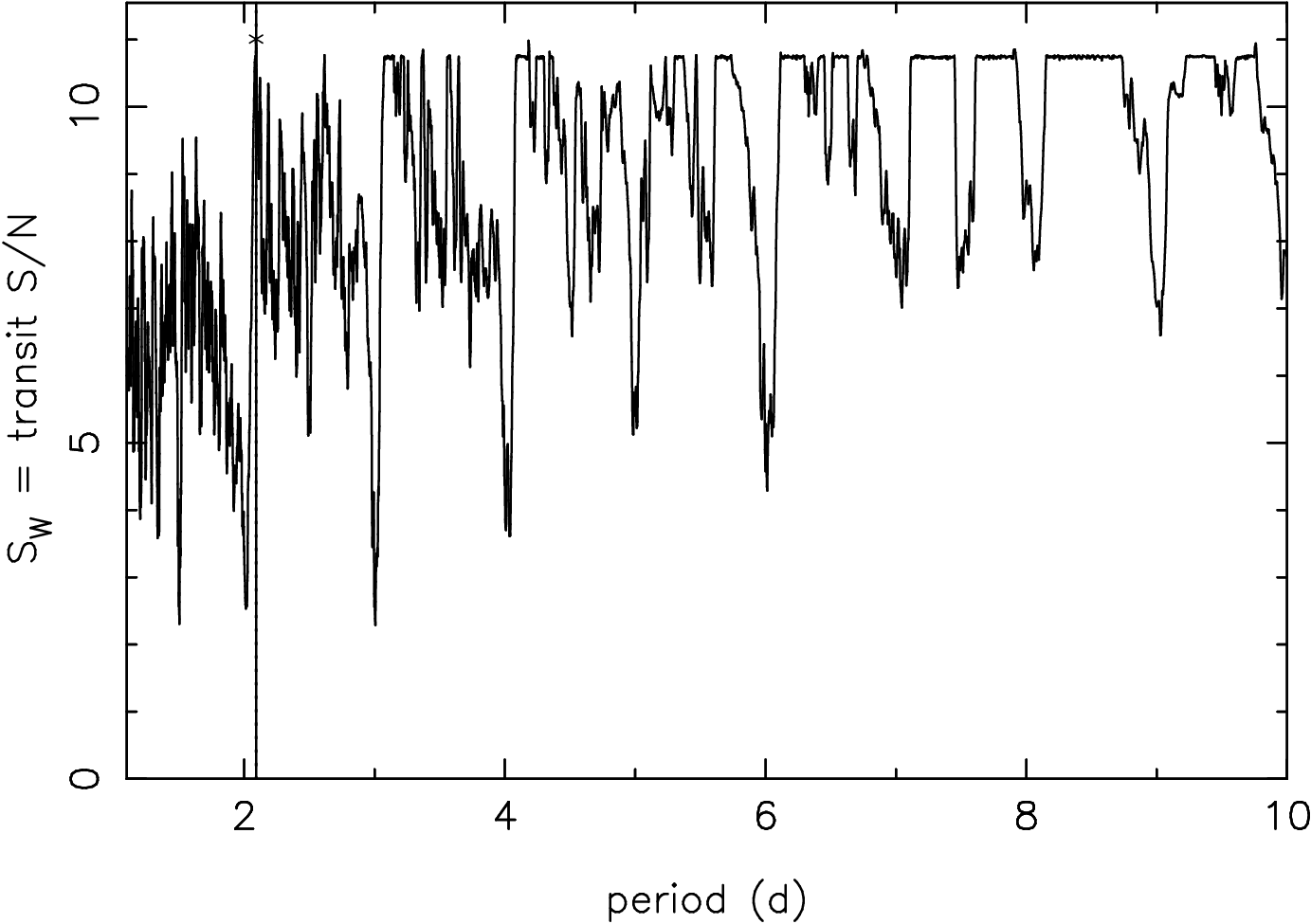} \\
    \end{tabular}
  \end{center}
  \caption{Transit candidates 11--14 (top to bottom) from the Draco
field.}
\end{figure*}

{\bf WASP0-TR-01:} Only one transit was observed and is missing the
egress. This creates a degeneracy for the duration measurement (see
associatied duration diagram). The colour and transit depth place this
candidate slightly outside the region for a 1.5 $R_J$ planet (see Figure
11). However, even if the egress had been observed, it is likely that the
duration and period don't match well for this to be considered a real
planetary transit.

{\bf WASP0-TR-02:} This appears to be a likely candidate with a strong
S/N and about 6 transits observed. Also, the duration and period are in
excellent agreement for a transit candidate. However, the colour of the
star shows that a 3.9 $R_J$ companion is required to produce the depth
fitted, and hence this is unlikely to be a real planetary transit.

{\bf WASP0-TR-03:} Around 2 transits were observed for this star and the
fitted duration and period match very well. The relatively large depth
though requires a very late-type star and indeed the colour results in a
4.0 $R_J$ companion estimate. Thus this cannot be considered a real
planetary transit.

{\bf WASP0-TR-04:} Although passing initial tests to be selected as a
transit candidate, this is unlikely to be real. Only one transit was
observed making the period uncertain. The egress was not observed making
the fitted duration also uncertain. The colour of the star and the large
depth result in a companion radius of 5.0 $R_J$.

{\bf WASP0-TR-05:} About 2 transits have been observed. However, the
duration matches poorly with the fitted period. With so few transits
observed, there are strong aliases at longer periods. The transit depth
and colour indicate a companion radius of 1.2 $R_J$. The transit is
``V-shaped'' which means that this is possibly due to a grazing
eclipsing binary rather than a true planetary transit.

{\bf WASP0-TR-06:} Only one transit was observed for this star casting
doubt upon the fitted period. The egress and bottom of the transit were
not observed thereby making the fitted duration and depth (which can be
treated as a lower limit) also highly uncertain. The colour and depth
imply a companion radius of 3.9 $R_J$. This is not considered to be a
real planetary transit.

{\bf WASP0-TR-07:} The rms scatter for this star causes the unbinned
lightcurve to appear very much like a planetary transit. Once binned
(10 minute bins) however, it becomes clear that this is in fact a low
amplitude variable star.

{\bf WASP0-TR-08:} Around 3 transits were observed for this star. The
duration and period match closely to what one would expect for a
planetary transit. The depth is relatively large and the associated star
colour results in an implied companion radius of 3.3 $R_J$. Hence this
is unlikely to be a real planetary transit.

{\bf WASP0-TR-09:} Though this was a strong candidate selected by the
transit detection algorithm, the binned lightcurve for this star clearly
reveals an eclipsing binary system. In particular, this binary appears
to be a magnetically active RSCVn binary with a period half that of
the fitted period, or around 0.73 days. The colour of the star indicates
an early K spectral type. According to the SIMBAD database \citep{wen00},
an x-ray counterpart for this source has been observed using ROSAT,
designated 1RXS J174211.8+465442.

{\bf WASP0-TR-10:} About 2 transits were observed for this candidate and
the duration matches well with the fitted period. However, the colour
and the depth imply that a 2.7 $R_J$ companion, making this unlikely to
be due to a planetary transit.

{\bf WASP0-TR-11:} The duration and period match well for a planetary
transit and many transits have been observed. However, binning the
lightcurve reveals secondary eclipses. This is in fact an EA/EB type
eclipsing binary with a period half that of the fitted period, or around
0.55 days. The colour of the star indicates an early K spectral type.

{\bf WASP0-TR-12:} This candidate has around 5 transits observed and has
a good match between the duration and period. It can be seen that the
real period is half that of the fitted period, or around 1.1 days. The
colour for this candidate implies a companion radius of 1.25 $R_J$
placing it well within planetary candidate range. An alternative
explanation for this lightcurve is an EA type eclipsing binary with
similar stellar radii producing dips of almost identical depth (such as
RX Her). This candidate is worth further follow-up.

{\bf WASP0-TR-13:} This lightcurve is a good example of a transit mimic
for which a strong S/N was calculated by the transit detection algorithm.
Although the duration and period are well matched, the transit itself is
too deep, the shape of the transit is distinctly ``V-shaped'', and there
is evidence of a secondary eclipse in the lightcurve, suggesting a
grazing eclipsing binary.

{\bf WASP0-TR-14:} The duration and period are well matched for this
candidate for which 2 transits were observed. However, the depth and
colour imply a companion radius of 3.3 $R_J$. This is therefore unlikely
to be a real planetary transit.

\section{Discussion}

The results of this study shed light on various issues relating to limits
on planetary abundances, transit search algorithms, and optimising
transit surveys. These will now be discussed in some detail.

\subsection{Limits on Planetary Companions around Field Stars}

Recent analysis of radial velocity surveys such as \citet{san03} has
shown that planets are preferentially found around stars with higher
metallicity. This conclusion is further strengthened by the null result
of the transit search in 47 Tucanae \citep{gil00}, an older population
with low metallicity. Transit searches in open clusters such as that
performed by \citet{bru03} also produced few candidates suggesting that
planetary ejection due to cluster dynamics may play a stronger role in
planetary abundances compared to planet formation around stars of
low/high metallicity. However, observations of the halo and surrounding
field stars of 47 Tucanae by \citet{wel04} also failed to detect any
transiting planets indicating that metallicity is indeed the dominant
factor. The field stars surveyed in Draco are predominantly G dwarfs in
the solar neighbourhood and therefore of solar metallicity.

Using the model described in section 4.1, the numbers of expected
transiting planets with periods in the range 1--10 days is $\sim 17$. Of
these, $\sim$ 86\% will transit multiple times during the course of the
Draco observations. Furthermore, only unblended stars with a high number
of epochs (40\% of the Draco stars) were searched for transits. Finally,
the S/N limits of fainter stars resulted in the transit detection
algorithm detecting 20\% of the injected transits. Thus, these
calculations lead to an expected number of $\sim 1.3$ observed transits in
the Draco dataset, or $\sim 1.2$ observed multiple transits.
It should of course be noted that the uncertainty in this calculation is
expected to be high since even a small change in one of the values will
have a significant impact on the result.

Two out of the 14 transit candidates presented in this paper,
WASP0-TR-05 and WASP0-TR-12, have the possibility of being due to real
transiting planets. However, WASP0-TR-05 has a substantial mis-match
between the fitted period and duration and so will not be considered
further. This detection rate is in remarkable agreeance with the
expected detection rate previously calculated and also with that
calculated by \citet{bro03}. The main contributors to the reduction of
our detection rate are blended stars and our photometric precision. If
one is able to overcome these two obstacles then one expects, for a
similar observing window and magnitude depth, to detect $\sim 15$
multiple transits in the Draco field. Assuming a Poisson distribution,
the probability of an event happening $x$ times is given by
\begin{equation}
  P(x) = \frac{\mu^x e^{-\mu}}{x!}
\end{equation}
where $\mu$ is the expected value. If WASP0-TR-12 is a real planetary
transit and is the only one in the field ($x = 1$), then the probability
is calculated to be $4.6 \times 10^{-6}$, or significant at the
3.6$\sigma$ level. If all the transit candidates are false and there are
none in the field ($x = 0$), then the probability is calculated to be
$3.1 \times 10^{-7}$, or significant at the 3.9$\sigma$ level.

Thus it would be surprising if a more sensitive survey of the Draco
field revealed no additional candidates since the metallicity of the
Draco stars is similar to that of the stars monitored by the radial
velocity surveys. There is no particular direction bias in which radial
velocity planets have been discovered since this method generally probes
a distance not substantially greater than the solar neighbourhood. A
hypothetical lack of planets in the Draco field may suggest that there is
indeed a constraint on the metallicity model of planetary abundances in
that direction. A lack of planets could also be due to the assumption
that planetary periods are approximately uniform in log space which is
possibly biased by radial velocity planet discoveries.

\subsection{False-Alarm Rate}

The false-alarm rate has been a major concern for transit-hunting teams
thus far and has vastly increased the necessary CPU and human-effort time
required to sift real transit events from the data. Of the 14 transit
candidates reported here, almost all appear to be variable stars of
some kind. According to the variable star catalogues available via
SIMBAD, these stars are all previously unknown variables. The exception
to this may be WASP0-TR-09, an RSCVn binary for which an x-ray
counterpart is known but has not previously been observed at optical
wavelengths. These variable stars produce a high transit S/N in the
transit detection algorithm leading to false positives. Assuming that
WASP0-TR-12 is a real planetary transit, the false-alarm rate of mimics
to planetary transits is 13:1.

A source of false alarms in the Draco dataset was encountered due to
the frames in the second half of the night being rotating by 180$^\circ$
relative to frames in the first half of the night. This was due to the
use of a German equatorial mount, as described in \citet{kan04}. The
effect of this was to change the shape of PSF profiles and thus cause
a magnitude shift in the lightcurves of blended stars. As previously
described, most of the blended stars were removed from the dataset prior
to analysis. Those blended stars which leaked through into the transit
detection algorithm were fitted with an integer day period. Since, as
shown in Figure 6, there is a low probability of observing integer day
periods, these stars were quickly identified and removed.

The transit detection algorithm could be improved by incorporating many
aspects of the transit selection criteria described in this paper.
Perhaps the easiest criterion to insert would be an approximate
calculation of the period/duration for each candidate and exclude it if
there is a significant mis-match. If colour information could be made
available, the depth could be translated into planetary radii, thus
excluding a major source of mimics. However, transit searches tend to be
computationally expensive algorithms and so the challenge is to reduce
the false-alarm rate whilst avoiding substantial increases in processing
time.

\subsection{Follow-up of Transit Candidates}

The announcement of various transit candidates have had a significant
impact on large telescope subscriptions for radial velocity follow-up.
This is particularly true for faint candidates, such as those announced
by the OGLE-III project \citep{uda02}. Optimal methods are required for
transit mimic elimination to avoid the unnecessary use of large telescope
time depending on the nature of the survey. In other words, it is
essential to make maximum use of the available photometry before
resorting to spectroscopic follow-up.

The first stage of following up transit candidates is to remove transit
mimics from the list. The major source of mimics is eclipsing binaries,
either grazing eclipsers of $\sim 1\%$ depth or blended eclipsers
contributing $\sim 1\%$ of light \citep{bro03}. In most cases,
particularly for wide-field surveys such as WASP0 where the stellar
profiles are heavily undersampled, straightforward multi-colour
observations using a 1.0m telescope can resolve many blended objects.
There are also smaller robotic telescopes available, such as Roboscope
\citep{hon00} and Sherlock \citep{kot04}, which can quickly perform
higher angular resolution, multi-colour photometry on transit candidates.

\section{Conclusions}

This paper describes observations of a field in Draco using the
Wide Angle Search for Planets prototype (WASP0). The observations took
place over a period of two months from La Palma during which 35000 stars
were monitored from this field. The data were reduced using a pipeline
which makes use of the Tycho-2 and USNO-B catalogues to provide an
astrometric solution for each frame. By considering fluxes measured using
multiple apertures, we are able to exclude blended stars from the sample
and thus improve our transit search.

By applying the noise model for the instrument and generating a
Besan\c{c}on model for the Draco field, fake data were generated which
match the magnitude distribution and epochs of the real data. The
metallicity and colours of the Besan\c{c}on model stars were used to
calculate the probability of each star harbouring a planet and the
stellar radii. Thus, planetary transits were randomly inserted into the
fake dataset. These simulated transits were used to test the transit
detection algorithm and to calculate the expected number transits in the
entire Draco dataset.

In total, 14000 stars were included in the transit search which yielded
14 transit candidates. Colours extracted from the 2MASS survey were used
to estimate the stellar radius and hence the companion radius for each
candidate. Of the 14 candidates, 2 were found to pass enough of the
selection criteria to be worth further follow-up. The remainder of the
candidates are variable stars including one RSCVn binary with an x-ray
counterpart. The false-alarm rate from this survey can be reduced in
future by incorporating some of the selection criteria into the detection
algorithm.

Transit searches have different selection effects from the radial
velocity surveys, finding shorter-period planets orbiting lower-mass
stars. This will determine how the planet abundance and short-period
cutoff seen in radial velocity surveys depend on stellar mass. The two
transit candidates identified in this survey are consistent with the
expected detection rate considering the constraints of the data and
observing window. If however there is indeed a significant lack of
transiting planets around Draco field stars then this would be
particularly surprising as the field is dominated by G dwarf stars with
solar metallicity, and therefore a relatively high probability of
harbouring planets. The location of the radial velocity planets do not
suggest any such directional bias for planet detection. Future surveys
with higher sensitivity, both ground-based and space-based, will
resolve this issue.

\section*{Acknowledgements}

The authors would like to thank PPARC for supporting this research and
the Nichol Trust for funding the WASP0 hardware.
This publication makes use of data products from the Two Micron All Sky
Survey, which is a joint project of the University of Massachusetts and
the Infrared Processing and Analysis Center/California Institute of
Technology, funded by the National Aeronautics and Space Administration
and the National Science Foundation.

\end{document}